\documentclass[12pt]{article}
\usepackage{amsmath}
\usepackage{floatflt}
\usepackage{amsfonts}
\usepackage{amssymb}
\usepackage{graphicx}
\usepackage{psfrag}
\usepackage{epsf}
\usepackage{slashed}
\usepackage{hyper}
\usepackage{cite}

\begin{document}

\title{Solution Transformations for GS String \\
in $AdS_5\times S^5$ by Conserved Quantities}
\author{Zhan-Yun Wang$^\mathrm{1}$ \thanks{
Email: zywang81@gmail.com},Xiao-Ning Xie$^\mathrm{2}$,Jun Feng$^\mathrm{1}$,
Yao-Xiong Wang$^\mathrm{1}$, \and Yu Zeng$^\mathrm{1}$ and Kang-Jie Shi$^%
\mathrm{1}$ \\
{\footnotesize {1)Institute of Modern Physics, Northwest University, Xi'an,
710069, China}} \\
{\footnotesize {2) SKLLQG, Institute of Earth Environment, Chinese Academy
of Sciences, Xi'an 710075, China}}}
\date{}
\maketitle

\begin{abstract}
For Light-cone gauge of Green-Schwarz superstring in $AdS_{5}\times S^{5}$
background, we fix two bosonic variables $x^{+}=\tau$ and $y^{9}=\sigma$,
and then perform the partial Legendre transformation of the remaining
bosonic variables. We then obtain a Lagrangian which is linear in velocity
after eliminating the metric of world sheet. For such a system, one can
formulate its poisson bracket and Hamiltonian. Since this system is free and
without constraint, the hierarchy of infinite nonlocal conserved quantities
given by Bena, Polchinski and Roiban, induce solution transformations due to
Jacobi identity.
\end{abstract}

\section{Introduction}

Because of AdS/CFT correspondence\cite{M97,GKP98,W98}, there has been much
interest in the role of integrability in the world-sheet theory of type IIB
strings in $AdS$ spaces. Metsaev and Tseytlin\cite{MT1} gave the famous $
\frac{PSU(2,2|4)}{SO(2,4)\otimes SO(5)}$ coset model with Wess-Zumino term
which describes string in $AdS_{5}\times S^{5}$ background. Because of $%
\kappa $ symmetry, the model has same degree of freedom for bosonic and
fermionic canonical variables. Its flat-space limit is the well known
Green-Schwarz superstring\cite{Green-Schwarz1,Green-Schwarz2}. This model
has attracted renewed interest and been studied in various aspects\cite{LU,
RR}. The parametrization is a hot issue of them, which includes the work by
Kallosh, Rajaraman, Rahmfeld(KRR)\cite{kallosh3}, Roiban and Siegel\cite{RS}%
, and many other authors\cite{Young,MT2, MT3, Pesando}. A Light-cone gauge
was given by Metsaev and Tseytlin using properly grouping $PSU(2,2|4)$ AdS
base \cite{MT2,MT3}. Another Light-cone gauge was given in $Z_{4}$\ grading
matrix approach by Alday, Arutyunov and Tseytlin, et al. with Hamiltonian
construction and quantization\cite{AFT, AF2 ,AF1}. These work simplifies the
Lagrangian of the model and investigates its solution and symmetry
properties.

After the construction of the Metsaev and Tseytlin's model, Bena, Polchinski
and Roiban\cite{BPR03}constructed one parameter flat currents which implies
a hierarchy of infinite nonlocal conserved quantities for the Green-Schwarz
superstring in $AdS_{5}\times S^{5}$ space-time. Thus the world-sheet sigma
model is probably completely integrable\cite{Vallilo}. This is a
breakthrough which attracts much attention and many corresponding studies in
string theory\cite{Zhou, Poly04, Hou, xiong, Chen05, Hatsuda04, Das04,
Kluson}. There are other approaches of the AdS strings, including their
quantization by Berkovits\cite{Berkovits1, Berkovits2, Berkovits3}.

For integrable models in two dimension field theory, the solution
transformation is a traditional topic. However, Metsaev and Tseytlin's model
is different from ordinary nonlinear $\sigma $-models in that it has a
Wess-Zumino term and satisfies the virasoro constraint. Consequently, we
can't use the usual methods such as Riemann-Hilbert transformation directly.

The dynamic structure and Hamiltonian of the model which are important for
the integrability were studied\cite{AFT,AF2,AF1,Kluson}. Our work is
motivated by these work, mostly by the Legendre transformation of bosonic
variables in the kinetic part of the Lagrangian originally introduced by
Alday, Arutyunov and Tseytlin, et al\cite{AFT}.

In this paper, we use the Light-cone $\kappa $ symmetry gauge by Metsaev and
Tseytlin\cite{MT2} and the $S^{5}$ parametrization by Kallosh, Rahmfeld,
Rajaraman and H. L\"{u}, et al\cite{LU, kallosh3}. We first fix $x^{+}=\tau $
and$\ y^{9}=\sigma $. After Legendre transformation of the remaining bosonic
variables in kinetic part of Lagrangian, the Lagrangian becomes linear in
'velocity' of canonical variables and is degenerate. This system is actually
a free Hamiltonian system without any constraint. The poisson bracket can be
induced from the final Lagrangian\cite{AFT,AF2,AF1}. We check that, the
Jacobi identity of the poisson bracket is satisfied for canonical variables,
Hamiltonian and conserved quantities given by Bena, et al\cite{BPR03}. Since
the poisson bracket of the conserved quantities and Hamiltonian must be
identically zero, then due to Jacobi identity one may generate solutions
from an existing one by these conserved quantities. The degrees of freedom
of such solution transformations is the same as the traditional
Riemann-Hilbert transformations.

This paper is organized as follows. In section 2 we briefly review the
Metsaev-Tseytlin formulation of superstring on the $AdS_{5}\times S^{5}$
background with $\kappa $ symmetry. In section 3, after the Light-cone $%
\kappa $ symmetry gauge fixing, we perform partial Legendre transform to
obtain the final degenerate Lagrangian which is linear in velocities of the
dynamic variables. In section 4, we study the poisson structure and prove
Jacobi identity for system of finite degree of freedom whose Lagrangian is
linear in velocities. The formal extension to field theory is also given. In
section 5, we firstly review the flat currents and the infinite conserved
quantities\ discovered by Bena, Polchinski and Roiban\cite{BPR03}. Then we
give the solution transformation from the conserved quantities. we make some
further discussions in the last section. Appendices include some detailed
calculations. In some parts of our paper, although the contents are known,
we present the explicit derivation for self-containment and for the
convenience to check the final results.

\section{Coset model of $PSU(2,2|4)$}

In this section, we first recall the superalgebra $psu(2,2|4)$\ and the
action of Green-Schwarz superstring in $AdS_{5}\times S^{5}$ spacetime
constructed by Metsaev and Tseytlin and its $\kappa $ symmetry\cite{MT1}\cite%
{me}.

Superstring propagating in the $AdS_{5}\times S^{5}$ spacetime can be
described as the non-linear sigma-model whose target space is the coset
superspace\cite{MT1}
\begin{equation*}
\frac{PSU(2,2|4)}{SO(4,1)\otimes SO(5)}
\end{equation*}
\ with the corresponding superalgebra $psu(2,2|4)$ in the $so(4|1)\oplus
so(5)$ basis.

\subsection{Superalgebra $psu(2,2|4)$}

The generators of $psu(2,2|4)$ are $T_{A}=(P_{a},J_{ab},P_{a^{\prime
}},J_{a^{\prime }b^{\prime }},Q_{\alpha \alpha ^{\prime }I})$, here the
indices $a,b,c,d=0,1,2,3,4$; $a^{\prime },b^{\prime },c^{\prime },d^{\prime
}=5,6,7,8,9$; $I,J=1,2$; $\alpha ,\beta =1,2,3,4$; $\alpha ^{\prime },\beta
^{\prime }=1,2,3,4$. The commutation relations for the generators of the Lie
superalgebra $psu(2,2|4)$ are%
\begin{equation*}
\lbrack P_{a},P_{b}]=J_{ab},\text{ \ \ \ \ \ \ \ \ \ \ \ \ \ \ \ \ \ \ \ \ \
\ \ \ \ \ \ \ \ \ }\left[ P_{a^{\prime }},P_{b^{\prime }}\right]
=-J_{a^{\prime }b^{\prime }},
\end{equation*}%
\begin{equation*}
\lbrack J_{ab},J_{cd}]=\eta _{bc}J_{ad}+\eta _{ad}J_{bc}-\eta
_{ac}J_{bd}-\eta _{bd}J_{ac},
\end{equation*}%
\begin{equation*}
\lbrack J_{a^{\prime }b^{\prime }},J_{c^{\prime }d^{\prime }}]=\eta
_{b^{\prime }c^{\prime }}J_{a^{\prime }d^{\prime }}+\eta _{a^{\prime
}d^{\prime }}J_{b^{\prime }c^{\prime }}-\eta _{a^{\prime }c^{\prime
}}J_{b^{\prime }d^{\prime }}-\eta _{b^{\prime }d^{\prime }}J_{a^{\prime
}c^{\prime }},
\end{equation*}%
\begin{equation*}
\lbrack P_{a},J_{bc}]=\eta _{ab}P_{c}-\eta _{ac}P_{b},\text{\ \ \ \ \ \ \ \
\ \ }[P_{a^{\prime }},J_{b^{\prime }c^{\prime }}]=\eta _{a^{\prime
}b^{\prime }}P_{c^{\prime }}-\eta _{a^{\prime }c^{\prime }}P_{b^{\prime }},
\end{equation*}%
\begin{equation*}
\lbrack Q_{\alpha \alpha ^{\prime }I},P_{a}]=-\frac{\mathrm{i}}{2}\epsilon
_{IJ}Q_{\beta \alpha ^{\prime }J}(\gamma _{a})_{\beta \alpha },\text{ \ \ \
\ \ }[Q_{\alpha \alpha ^{\prime }I},P_{a^{\prime }}]=\frac{1}{2}\epsilon
_{IJ}Q_{\alpha \beta ^{\prime }J}(\gamma _{a^{\prime }})_{\beta ^{\prime
}\alpha ^{\prime }},
\end{equation*}%
\begin{equation*}
\lbrack Q_{\alpha \alpha ^{\prime }I},J_{ab}]=-\frac{1}{2}Q_{\beta \alpha
^{\prime }I}(\gamma _{ab})_{\beta \alpha },\text{ \ \ \ \ \ \ }[Q_{\alpha
\alpha ^{\prime }I},J_{a^{\prime }b^{\prime }}]=-\frac{1}{2}Q_{\alpha \beta
^{\prime }I}(\gamma _{a^{\prime }b^{\prime }})_{\beta ^{\prime }\alpha
^{\prime }},
\end{equation*}%
\begin{align}
& \{Q_{\alpha \alpha ^{\prime }I},Q_{\beta \beta ^{\prime }J}\}  \notag \\
& =\delta _{IJ}\left[ -2\mathrm{i}C_{\alpha ^{\prime }\beta ^{\prime
}}^{\prime }(C\gamma ^{a})_{\alpha \beta }P_{a}+2C_{\alpha \beta }(C^{\prime
}\gamma ^{a^{\prime }})_{\alpha ^{\prime }\beta ^{\prime }}P_{a^{\prime }}%
\right]  \notag \\
& +\epsilon _{IJ}\left[ C_{\alpha ^{\prime }\beta ^{\prime }}^{\prime
}(C\gamma ^{ab})_{\alpha \beta }J_{ab}-C_{\alpha \beta }(C^{\prime }\gamma
^{a^{\prime }b^{\prime }})_{\alpha ^{\prime }\beta ^{\prime }}J_{a^{\prime
}b^{\prime }}\right],
\end{align}%
where $\eta _{ab}=(-++++)$, $\eta _{a^{\prime }b^{\prime }}=(+++++),\left\{
\gamma ^{a},\gamma ^{b}\right\} =2\eta ^{ab},\left\{ \gamma ^{a^{\prime
}},\gamma ^{b^{\prime }}\right\} =2\eta ^{a^{\prime }b^{\prime }},\epsilon
_{12}=-\epsilon _{21}=1.$ Here and after, the repeated indices are summed.

The gamma matrices satisfy
\begin{eqnarray*}
\left\{ \gamma ^{a},\gamma ^{b}\right\} &=&2\eta ^{ab},\left\{ \gamma
^{a^{\prime }},\gamma ^{b^{\prime }}\right\} =2\eta ^{a^{\prime }b^{\prime
}}, \\
\gamma ^{ab} &=&\frac{1}{2}[\gamma ^{a},\gamma ^{b}],\gamma ^{a^{\prime
}b^{\prime }}=\frac{1}{2}[\gamma ^{a},\gamma ^{b}], \\
(\mathcal{C}\gamma ^{\hat{a}})^{t} &=&\mathcal{C}\gamma ^{\hat{a}},(\mathcal{%
C}\gamma ^{\hat{a}\hat{b}})^{t}=-\mathcal{C}\gamma ^{\hat{a}\hat{b}},%
\mathcal{C}^{t}=\mathcal{C}=C\otimes C^{\prime },\hat{a},\hat{b}=a,b\text{ \
}\text{or}\text{ \ }a^{\prime },b^{\prime }.
\end{eqnarray*}%
We may set
\begin{equation*}
\gamma _{a}=\left[
\begin{array}{cc}
0 & \sigma _{a} \\
\bar{\sigma}_{a} & 0%
\end{array}%
\right],\sigma _{a}=(\mathbf{I},\sigma _{1},\sigma _{2},\sigma _{3}),\bar{
\sigma}_{a}=(-\mathbf{I},\sigma _{1},\sigma _{2},\sigma _{3}),
\end{equation*}%
for $a=0,1,2,3,$ where $\mathbf{I}$ is the $2\times 2$ unit matrix,
\begin{equation*}
\gamma _{4}=\left[
\begin{array}{cc}
\mathbf{I} & 0 \\
0 & -\mathbf{I}%
\end{array}
\right],\gamma _{5}=\text{i}\gamma _{0},\gamma _{6}=\gamma _{1},\gamma
_{7}=\gamma _{2},\gamma _{8}=\gamma _{3},\gamma _{9}=\gamma _{4},
\end{equation*}%
and
\begin{equation}
C=C^{\prime }=\left[
\begin{array}{cc}
i\sigma ^{2} & 0 \\
0 & -i\sigma ^{2}%
\end{array}%
\right] .  \label{c}
\end{equation}

The left-invariant Cartan 1-forms
\begin{equation}
L^{A}=dX^{M}L_{M}^{A},\quad X^{M}=(x,\theta ),
\end{equation}%
are given by
\begin{equation}
\mathcal{J}=G^{-1}dG=L^{A}T_{A}\equiv L^{a}P_{a}+L^{a^{\prime }}P_{a^{\prime
}}+\frac{1}{2}L^{ab}J_{ab}+\frac{1}{2}L^{a^{\prime }b^{\prime }}J_{a^{\prime
}b^{\prime }}+L^{\alpha \alpha ^{\prime }I}Q_{\alpha \alpha ^{\prime }I}\ ,
\label{cartanform}
\end{equation}%
where $G=G(x,\theta )$ is a coset representative in $PSU(2,2|4)$.

The Cartan 1-form satisfies the zero-curvature equation $d\mathcal{J}=-
\mathcal{J}\wedge \mathcal{J}$. When decompose it according to the
generators of the Lie algebra, we\ get Maurer-Cartan equations.

\subsection{$\protect\kappa $ symmetry}

The string theory action is the sum of the non-linear sigma-model action and
a topological Wess-Zumino term to ensure $\kappa $ symmetry. The Polyakov
action given by Metsaev and Tseytlin\cite{MT1} is%
\begin{eqnarray}
S &=&S_{k}+S_{WZ}  \notag  \label{action} \\
&=&-\frac{1}{2}\int_{\partial M_{3}}d^{2}\sigma \sqrt{-g}%
g^{ij}(L_{i}^{a}L_{j}^{a}+L_{i}^{a^{\prime }}L_{j}^{a^{\prime
}})-\int_{\partial M_{3}}d^{2}\sigma \epsilon ^{ij}(\bar{L}_{i}^{1}L_{j}^{2}+%
\bar{L}_{i}^{2}L_{j}^{1}).  \label{in}
\end{eqnarray}%
Here $\sqrt{-g}=\sqrt{-\det g_{ij}},g_{ij}g^{jk}=\delta _{ik},i,j=0,1$ and $%
g_{ij}$\ is the metric of the world-sheet. $2\times 2$ matrix $\epsilon
^{ij}=-\epsilon ^{ji},\ \epsilon ^{01}=1.$ Here and after denote $X^{\hat{a}%
}Y^{\hat{a}}=X^{\hat{a}}Y^{\hat{b}}\eta _{\hat{a}\hat{b}}$. This action is
invariant with respect to the local $\kappa $-transformations. Due to $\bar{L%
}=L^{\dag }\gamma _{0}=L^{t}CC^{\prime }$ for Majorana spinor $L$, the WZ
term can be written as $S_{WZ}=-\int_{\partial M_{3}}d^{2}\sigma \epsilon
^{ij}(L_{i}^{1t}CC^{\prime }L_{j}^{2}+L_{i}^{2t}CC^{\prime }L_{j}^{1}).$

Let the variation of group element be $\delta G$ and $\rho =G^{-1}\delta
G\equiv P_{a}\delta x^{a}+P_{a^{\prime }}\,\delta x^{a^{\prime }}+\frac{1}{2}%
J_{ab}\,\delta x^{ab}+\frac{1}{2}J_{a^{\prime }b^{\prime }}\,\delta
x^{a^{\prime }b^{\prime }}+Q_{\alpha \alpha ^{\prime }}^{I}\delta \theta
^{I\alpha \alpha ^{\prime }}$. The equation $\delta \mathcal{J}=d\rho +[%
\mathcal{J},\rho ]$ gives the variations of the Cartan 1-forms. The
variation of action (\ref{in}) with respect to $\delta x^{a},\delta
x^{a^{\prime }},\delta \theta ^{I\alpha \alpha ^{\prime }}$ gives the
equations of motion
\begin{eqnarray}
\partial _{i}(\gamma ^{ij}L_{j}^{a})+\gamma ^{ij}L_{i}^{ab}L_{j}^{b}+\mathrm{%
i}\epsilon ^{ij}s^{IJ}\bar{L}_{i}^{I}\gamma ^{a}L_{j}^{J} &=&0,  \notag \\
\partial _{i}(\gamma ^{ij}L_{j}^{a^{\prime }})+\gamma ^{ij}L_{i}^{a^{\prime
}b^{\prime }}L_{j}^{b^{\prime }}-\epsilon ^{ij}s^{IJ}\bar{L}_{i}^{I}\gamma
^{a^{\prime }}L_{j}^{J} &=&0,  \notag \\
(L_{i}^{a}\gamma ^{a}+iL_{i}^{a^{\prime }}\gamma ^{a^{\prime }})(\gamma
^{ij}-\epsilon ^{ij})L_{j}^{1} &=&0,\   \notag \\
(L_{i}^{a}\gamma ^{a}+iL_{i}^{a^{\prime }}\gamma ^{a^{\prime }})(\gamma
^{ij}+\epsilon ^{ij})L_{j}^{2} &=&0.  \label{EM}
\end{eqnarray}%
while the variation of the metric $g_{ij}$\ gives the virasoro constraint
\begin{equation}
L_{i}^{a}L_{j}^{a}+L_{i}^{a^{\prime }}L_{j}^{a^{\prime }}=\frac{1}{2}%
g_{ij}g^{kl}\left( L_{k}^{a}L_{l}^{a}+L_{k}^{a^{\prime }}L_{l}^{a^{\prime
}}\right) ,  \label{virasoro}
\end{equation}%
where $\gamma ^{ij}=\sqrt{-g}g^{ij}$, $\det [\gamma _{ij}]=-1.$ $
s^{IJ}=(\sigma _{3})^{IJ}=(-1)^{I-1}\delta _{IJ}$.

Substituting the virasoro constraint(\ref{virasoro}) into (\ref{action}),
one obtains the Nambu-Goto action%
\begin{equation}
S=\mathfrak{-}\int_{\partial M_{3}}(d^{2}\sigma \sqrt{-\mathcal{G}}+2\bar{L}%
^{1}\wedge L^{2}),
\end{equation}%
where the induced metric $\mathcal{G}_{ij}=L_{i}^{a}L_{j}^{a}+L_{i}^{a^{
\prime }}L_{i}^{a^{\prime }}$ and $\mathcal{G}=\det [\mathcal{G}_{ij}]$.

We may check the $\kappa $ symmetry in the Nambu-Goto action, which can give
the right degrees of freedom\cite{me}. Consider the variation of $\delta
_{\kappa }\bar{\theta}^{I}$, one obtains

\begin{equation}
\delta _{\kappa }S\mathcal{=}4\text{i}\int_{\partial M_{3}}d^{2}\sigma \sqrt{%
-\mathcal{G}}(\delta _{\kappa }\bar{\theta}^{1}P_{+}^{ij}\slashed{L}%
_{j}^{+}L_{i}^{1}+\delta _{\kappa }\bar{\theta}^{2}P_{-}^{ij}\slashed{L}%
_{j}^{+}L_{i}^{2}),
\end{equation}%
where

\begin{equation}
\slashed{L}_{i}^{+}\equiv (L_{i}^{a}\gamma ^{a}+iL_{i}^{a^{\prime }}\gamma
^{a^{\prime }}),\text{ \ }\slashed{L}_{i}^{-}\equiv (L_{i}^{a}\gamma
^{a}-iL_{i}^{a^{\prime }}\gamma ^{a^{\prime }}).
\end{equation}

Define
\begin{eqnarray}
P_{\pm }^{ij} &=&\frac{1}{2}(\mathcal{G}^{ij}\pm \frac{\epsilon ^{ij}}{\sqrt{%
-\mathcal{G}}}),  \notag \\
\gamma &=&-\frac{\epsilon ^{ij}\slashed{L}_{i}^{+}\slashed{L}_{j}^{-}}{2%
\sqrt{-\mathcal{G}}},\gamma ^{2}=1,\mathbf{tr}\gamma =0.  \label{gama}
\end{eqnarray}

One has
\begin{equation}
\gamma P_{\pm }^{ij}\slashed{L}_{j}^{+}=\pm P_{\pm }^{ij}\slashed{L}_{j}^{+},
\label{11a}
\end{equation}%
\begin{equation}
(1\mp \gamma )P_{\pm }^{ij}\slashed{L}_{j}^{+}=0.  \label{11aa}
\end{equation}%
The local $\kappa $-transformations can be written as
\begin{align}
\delta _{\kappa }x^{\hat{a}}& =\delta _{\kappa }x^{\hat{a}\hat{b}}=0,  \notag
\\
\delta _{\kappa }\bar{\theta}^{1}& =\bar{\kappa}^{1}P_{-},  \notag \\
\delta _{\kappa }\bar{\theta}^{2}& =\bar{\kappa}^{2}P_{+},  \label{k va}
\end{align}%
where $\bar{\kappa}^{1}$and $\bar{\kappa}^{2}$ are arbitrary, $P_{\pm }=%
\frac{1\pm \gamma }{2}$ are projector operators. For such variation, we have
$\delta _{\kappa }S=0$.

This is a local symmetry of the model, thus this system is not definite,
which has infinite solutions for given initial and boundary conditions. We
must perform $\kappa $ symmetry gauge fixing, only take half of the
fermionic variables.

\section{Light-cone gauge fixing and partial Legendre transformation}

\subsection{parametrization}

In this subsection, we follow the Light-cone $\kappa $ symmetry gauge fixing
by Metsaev and Tseytlin\cite{MT2}, while the $S^{5}$ part we use the
parametrization of KRR\cite{kallosh3}, and H. L\"{u} et al\cite{LU}.

Define
\begin{eqnarray}
x^{\pm } &=&\frac{1}{\sqrt{2}}(x^{3}\pm x^{0}),\text{ \ \ \ \ }x=\frac{1}{%
\sqrt{2}}(x^{1}+\mathrm{i}x^{2}),  \notag \\
\bar{x} &=&\frac{1}{\sqrt{2}}(x^{1}-\mathrm{i}x^{2}),\text{ \ \ \ \ }\phi
=x^{4},  \notag \\
x^{a} &=&(x^{+},x^{-},x,\bar{x},\phi ),\text{ \ \ }\eta ^{+-}=\eta
^{-+}=\eta ^{x\bar{x}}=\eta ^{\bar{x}x}=1.  \label{3-a}
\end{eqnarray}%
The bosonic\ generators of AdS for Light-cone gauge are
\begin{eqnarray*}
D &=&P^{4}, \\
P^{\pm } &=&\frac{1}{\sqrt{2}}(P^{3}\pm P^{0}+J^{43}\pm J^{40});\text{ \ \ \
}P^{x/\bar{x}}=\frac{1}{\sqrt{2}}(P^{1}\pm \mathrm{i}P^{2}+J^{41}\pm \mathrm{%
i}J^{42}); \\
K^{\pm } &=&\frac{1}{2\sqrt{2}}(-P^{3}\mp P^{0}+J^{43}\pm J^{40});\text{ \ }%
K^{x/\bar{x}}=\frac{1}{2\sqrt{2}}(-P^{1}\mp \mathrm{i}P^{2}+J^{41}\pm
\mathrm{i}J^{42});
\end{eqnarray*}%
\begin{eqnarray}
J^{\pm x} &=&\pm \frac{1}{2}J^{01}\pm \frac{\mathrm{i}}{2}J^{02}+\frac{1}{2}%
J^{31}+\frac{\mathrm{i}}{2}J^{32};  \notag \\
J^{\pm \bar{x}} &=&\pm \frac{1}{2}J^{01}\mp \frac{\mathrm{i}}{2}J^{02}+\frac{%
1}{2}J^{31}-\frac{\mathrm{i}}{2}J^{32};  \notag \\
J^{x\bar{x}} &=&-\mathrm{i}J^{12};\text{ \ \ \ \ \ \ \ \ \ \ \ \ \ \ \ \ \ }%
J^{+-}=J^{03}.  \label{3-b}
\end{eqnarray}

Also define for fermionic generators
\begin{equation*}
Q_{\pm \alpha \alpha ^{\prime }}=Q_{1\alpha \alpha ^{\prime }}\pm \mathrm{i}%
Q_{2\alpha \alpha ^{\prime }},\text{ \ \ }q_{\alpha \alpha ^{\prime
}}=Q_{-\alpha \alpha ^{\prime }},\text{ \ \ }q^{\alpha \alpha ^{\prime
}}=C_{\alpha \beta }C_{\alpha ^{\prime }\beta ^{\prime }}^{\prime }Q_{+\beta
\beta ^{\prime }},
\end{equation*}%
and
\begin{eqnarray}
q^{1i} &=&\mathrm{i}{\normalsize 2}\sqrt{2}2^{-\frac{1}{4}}Q^{-i};\text{ }%
q^{2i}=-\mathrm{i}{\normalsize 2}\sqrt{2}2^{-\frac{1}{4}}Q^{+i};  \notag \\
q^{3i} &=&{\normalsize 2}\sqrt{2}2^{\frac{1}{4}}S^{+i};\text{ \ \ \ }q^{4i}=%
{\normalsize 2}\sqrt{2}2^{\frac{1}{4}}S^{-i};  \notag \\
q_{1i} &=&{\normalsize 2}\sqrt{2}2^{\frac{1}{4}}S_{i}^{+};\text{ \ \ \ \ }%
q_{2i}={\normalsize 2}\sqrt{2}2^{\frac{1}{4}}S_{i}^{-};  \notag \\
q_{3i} &=&\mathrm{i}{\normalsize 2}\sqrt{2}2^{-\frac{1}{4}}Q_{i}^{-};\text{\
\ }q_{4i}=-\mathrm{i}{\normalsize 2}\sqrt{2}2^{-\frac{1}{4}}Q_{i}^{+},
\label{3-c}
\end{eqnarray}%
where the index $i$ is the $S^{5}$ index $\alpha ^{\prime }$.

We then take the parametrization following\cite{MT2}, and \cite{kallosh3, LU}
for $S^{5}$ part. The $\kappa $-symmetry gauge fixed representative group
element is
\begin{eqnarray}
G(x,y,\theta ,\eta ,\phi ) &=&g\left( x\right) g\left( \theta \right)
g\left( \eta \right) g(y)g(\phi )  \notag  \label{coset} \\
g\left( x\right) &=&\exp (x^{-}P^{+}+x^{+}P^{-}+xP^{\bar{x}}+\bar{x}P^{x});
\notag \\
g\left( \theta \right) &=&\exp (\theta ^{i}Q_{i}^{+}+\theta _{i}Q^{+i});
\notag \\
g\left( \eta \right) &=&\exp (\eta ^{i}S_{i}^{+}+\eta _{i}S^{+i});  \notag \\
g(y)
&=&e^{y^{5}J_{56}}e^{y^{6}J_{67}}e^{y^{7}J_{78}}e^{y^{8}J_{89}}e^{y^{9}P_{9}};
\notag \\
g(\phi ) &=&\exp (\phi D).
\end{eqnarray}%
Define
\begin{equation*}
M=\exp (\frac{y_{5}}{2}\gamma ^{56})\exp (\frac{y_{6}}{2}\gamma ^{67})\exp (%
\frac{y_{7}}{2}\gamma ^{78})\exp (\frac{y_{8}}{2}\gamma ^{89})\exp (-\frac{%
{\normalsize i}y_{9}}{2}\gamma ^{9}),
\end{equation*}%
and
\begin{equation*}
\tilde{\xi}_{j}=\xi _{i}M_{ij},\tilde{\xi}^{i}=M_{ij}^{-1}\xi ^{j},
\end{equation*}%
for fermionic variables.

The one form of G is
\begin{align*}
\mathcal{J}& =G^{-1}\left( x,y,\theta ,\eta ,\phi \right) dG\left(
x,y,\theta ,\eta ,\phi \right) \\
& =L_{p}^{\pm }P^{\mp }+L_{p}^{x}P^{\bar{x}}+L_{p}^{\bar{x}%
}P^{x}+L^{D}D+L_{K}^{\pm }K^{\mp }+L_{K}^{x}K^{\bar{x}}+L_{K}^{\bar{x}%
}K^{x}+\,L^{a^{\prime }}P^{a^{\prime }} \\
& +L^{\mp x}J^{\pm \bar{x}}+L^{\pm \bar{x}}J^{\mp x}+L^{\bar{x}x}J^{x\bar{x}%
}+L^{+-}J^{-+} \\
& \,+\frac{1}{2}L^{a^{\prime }b^{\prime }}J_{a^{\prime }b^{\prime
}}\,+L_{Q}^{\pm i}Q_{i}^{\mp }+L_{S}^{\pm i}S_{i}^{\mp }+L_{iQ}^{\pm }Q^{\mp
i}+L_{iS}^{\pm }S^{\mp i}.
\end{align*}%
By the deliberately designed coset representative(\ref{coset}), one obtains
the nonzero 1-forms
\begin{eqnarray}
L^{A^{\prime }} &=&(\prod\limits_{k^{\prime }=A^{\prime }+1}^{9}\sin
y_{k^{\prime }}){\normalsize d}y_{A^{\prime }}-\frac{1}{2}{\normalsize d}%
x^{+}\tilde{\eta}_{i}(\gamma ^{A^{\prime }})_{j}^{i}\tilde{\eta}^{j}\equiv
u^{A^{\prime }}{\normalsize d}y_{A^{\prime }}+v^{A^{\prime }}{\normalsize d}%
x^{+},(u^{9}=1)  \notag \\
L^{A^{\prime }B^{\prime }} &=&-\frac{\mathrm{i}}{2}{\normalsize d}x^{+}%
\tilde{\eta}_{i}(\gamma ^{A^{\prime }B^{\prime }})_{j}^{i}\tilde{\eta}%
^{j}+\prod\limits_{k^{\prime }=A^{\prime }+1}^{B^{\prime }-1}\sin
y^{k^{\prime }}\cos y^{B^{\prime }}{\normalsize d}y^{A^{\prime }}  \notag \\
L_{p}^{+} &=&e^{\phi }{\normalsize d}x^{+},\text{ \ \ \ \ \ \ \ \ \ }%
L_{p}^{-}=e^{\phi }[{\normalsize d}x^{-}-\frac{\mathrm{i}}{2}(\theta _{i}%
{\normalsize d}\theta ^{i}+\theta ^{i}{\normalsize d}\theta _{i})],  \notag
\\
L_{p}^{\bar{x}} &=&e^{\phi }{\normalsize d}\bar{x},\text{ \ \ \ \ \ \ \ \ \
\ }L_{p}^{x}=e^{\phi }{\normalsize d}x,\text{ \ \ \ \ \ \ \ \ \ \ \ }%
L^{D}=d\phi ,  \notag \\
\text{\ \ \ }L_{K}^{-} &=&e^{-\phi }[\frac{1}{4}{\normalsize d}x^{+}({\eta
^{i}\eta _{i}})^{2}+\frac{\mathrm{i}}{2}(\eta _{i}{\normalsize d}\eta
^{i}+\eta ^{i}{\normalsize d}\eta _{i})],\text{ \ \ }L^{\bar{x}x}=\frac{%
\mathrm{i}}{2}dx^{+}(\eta ^{i}\eta _{i}),  \notag \\
L^{-x} &=&\eta _{i}{\normalsize d}\theta ^{i}-\frac{1}{2}dx(\eta ^{i}\eta
_{i}),\text{ \ \ \ \ \ \ \ \ \ \ }L^{-\bar{x}}=-\eta ^{i}{\normalsize d}%
\theta _{i}+\frac{\mathrm{i}}{2}d\bar{x}(\eta ^{i}\eta _{i}),  \notag \\
L_{Q}^{-i} &=&e^{\frac{1}{2}\phi }(\widetilde{{\normalsize d}\theta }^{i}+%
\mathrm{i}{\normalsize d}x\tilde{\eta}^{i}),\text{ \ \ \ \ \ \ \ \ \ \ \ \ }%
L_{Q}^{+i}=-\mathrm{i}e^{\frac{1}{2}\phi }{\normalsize d}x^{+}\tilde{\eta}%
^{i},  \notag \\
L_{iQ}^{-} &=&e^{\frac{1}{2}\phi }(\widetilde{{\normalsize d}\theta }_{i}-%
\mathrm{i}{\normalsize d}\bar{x}\tilde{\eta}_{i}),\text{ \ \ \ \ \ \ \ \ \ \
\ \ }L_{iQ}^{+}=\mathrm{i}e^{\frac{1}{2}\phi }{\normalsize d}x^{+}\tilde{\eta%
}_{i},  \notag \\
L_{S}^{-i} &=&e^{-\frac{1}{2}\phi }({\frac{\mathrm{i}}{2}}dx^{+}(\eta
^{k}\eta _{k})\tilde{\eta}^{i}+\tilde{d\eta ^{i}}),\text{ \ }L_{iS}^{-}=e^{-%
\frac{1}{2}\phi }(-{\frac{\mathrm{i}}{2}}dx^{+}(\eta ^{k}\eta _{k})\tilde{%
\eta}_{i}+\tilde{d\eta _{i}}).  \label{8-20}
\end{eqnarray}

Using the formula $L^{a}\equiv L_{p}^{a}-{\frac{1}{2}}L_{k}^{a}$ one gives
\begin{equation}
L^{-}=e^{\phi }[{\normalsize d}x^{-}-\frac{\mathrm{i}}{2}(\theta _{i}%
{\normalsize d}\theta ^{i}+\theta ^{i}{\normalsize d}\theta _{i})]-\frac{1}{2%
}e^{-\phi }[\frac{1}{4}{\normalsize d}x^{+}({\eta ^{i}\eta _{i}})^{2}+\frac{%
\mathrm{i}}{2}(\eta _{i}{\normalsize d}\eta ^{i}+\eta ^{i}{\normalsize d}%
\eta _{i})],  \label{3-3}
\end{equation}%
and has
\begin{eqnarray*}
\mathcal{L}_{k} &=&-{\frac{1}{2}}\sqrt{-g}g^{\mu \nu }(L_{\mu }^{+}L_{\nu
}^{-}+L_{\mu }^{-}L_{\nu }^{+}+L_{\mu }^{x}L_{\nu }^{{\bar{x}}}+L_{\mu }^{{%
\bar{x}}}L_{\nu }^{x}+L_{\mu }^{D}L_{\nu }^{D}+\sum_{A^{\prime
}=5}^{9}L_{\mu }^{A^{\prime }}L_{\nu }^{A^{\prime }}) \\
&\equiv &-{\frac{1}{2}}\gamma ^{\mu \nu }(x_{\mu }^{\hat{a}}G_{\hat{a}\hat{b}%
}x_{\nu }^{\hat{b}}) \\
&=&-{\frac{1}{2}}\gamma ^{\mu \nu }\{e^{\phi }[\partial _{\mu }x^{-}-{\frac{i%
}{2}}(\theta ^{i}\partial _{\mu }\theta _{i}+\theta _{i}\partial _{\mu
}\theta ^{i})]e^{\phi }\partial _{\nu }x^{+} \\
&&-\frac{1}{2}e^{-\phi }[\frac{1}{4}\partial _{\mu }x^{+}({\eta ^{i}\eta _{i}%
})^{2}+\frac{\mathrm{i}}{2}(\eta _{i}\partial _{\mu }\eta ^{i}+\eta
^{i}\partial _{\mu }\eta _{i})]e^{\phi }\partial _{\nu }x^{+} \\
+\mu &\leftrightarrow &\nu \\
&&+e^{2\phi }(\partial _{\mu }\bar{x}\partial _{\nu }x+\partial _{\mu
}x\partial _{\nu }\bar{x})+\partial _{\mu }\phi \partial _{\nu }\phi \\
&&+\sum_{A^{\prime }=5}^{9}(u^{A^{\prime }}\partial _{\mu }y_{A^{\prime
}}+v^{A^{\prime }}\partial _{\mu }x^{+})(u^{A^{\prime }}\partial _{\nu
}y_{A^{\prime }}+v^{A^{\prime }}\partial _{\nu }x^{+})\}
\end{eqnarray*}%
\begin{eqnarray}
&=&\gamma ^{\mu \nu }\{-e^{2\phi }(\partial _{\mu }x^{+}\partial _{\nu
}x^{-}+\partial _{\mu }x\partial _{\nu }\bar{x})-{\frac{1}{2}}\partial _{\mu
}\phi \partial _{\nu }\phi  \notag \\
&&-{\frac{1}{2}}\sum_{A^{\prime }=5}^{9}(u^{A^{\prime }}\partial _{\mu
}y^{A^{\prime }})(u^{A^{\prime }}\partial _{\nu }y^{A^{\prime }})+\partial
_{\mu }x^{+}[{\frac{i}{2}}e^{2\phi }(\theta ^{i}\partial _{\nu }\theta
_{i}+\theta _{i}\partial _{\nu }\theta ^{i})  \notag \\
&&+{\frac{i}{4}}(\eta ^{i}\partial _{\nu }\eta _{i}+\eta _{i}\partial _{\nu
}\eta )-\sum_{A^{\prime }=5}^{9}v^{A^{\prime }}u^{A^{\prime }}\partial _{\nu
}y^{A^{\prime }}]  \notag \\
&&+{\frac{1}{8}}\partial _{\mu }x^{+}\partial _{\nu }x^{+}[(\eta ^{i}\eta
_{i})^{2}-4\sum_{A^{\prime }=5}^{9}(v{^{A^{\prime }}})^{2}]\},  \label{3-f}
\end{eqnarray}%
with%
\begin{eqnarray}
G_{+-} &=&G_{-+}=1,G_{x\bar{x}}=G_{\bar{x}x}=1,G_{DD}=G_{{A^{\prime
}A^{\prime }}}=1,A^{\prime }=5,\cdots 9,  \notag \\
\text{and \ }\gamma ^{\mu \nu } &=&\sqrt{-g}g^{\mu \nu },x_{\mu }^{\hat{a}%
}=L_{\mu }^{\hat{a}}.  \label{23a}
\end{eqnarray}

We also have
\begin{eqnarray}
\mathcal{L}_{WZ} &=&-\epsilon ^{\mu \nu }(L_{\mu }^{1}CC^{\prime }L_{\nu
}^{2}+L_{\mu }^{2}CC^{\prime }L_{\nu }^{1})  \notag \\
&=&-{\frac{\mathrm{i}\epsilon ^{\mu \nu }}{\sqrt{2}}}\{\mathrm{i}e^{\phi
}\partial _{\mu }x^{+}\tilde{\eta _{i}}C_{ij}^{\prime }(\tilde{\partial
_{\nu }\theta _{j}}-\mathrm{i}\partial _{\nu }\bar{x}\tilde{\eta _{j}})-%
\mathrm{i}e^{\phi }\partial _{\mu }x^{+}\tilde{\eta ^{i}}C_{ij}^{\prime }(%
\tilde{\partial _{\nu }\theta ^{j}}+\mathrm{i}\partial _{\nu }x\tilde{\eta
^{j}})\}.
\end{eqnarray}

\subsection{Partial Legendre transformation}

We next fix the gauge $x^{+}=\tau ,y^{9}=\sigma $ and perform partial
Legendre transformation for the remaining bosonic variables in $\mathcal{L}%
_{k}$.

Define ${\frac{\partial \mathcal{L}_{k}}{\partial \dot{z}^{i}}}=\pi _{i}$
for 8 bosonic variables. Perform Legendre transformation partially, we
have(Appendix A)
\begin{eqnarray}
\mathcal{\tilde{H}} &=&\pi _{i}\dot{z}_{i}-\mathcal{L}_{k}  \notag \\
&=&{\frac{e^{2\phi }}{{2\pi _{-}}}}\{2e^{2\phi }x^{\prime }\bar{x}^{\prime }+%
{\phi ^{\prime }}^{2}+\sum_{a^{\prime }=5}^{8}(u^{a^{\prime
}})^{2}(y^{\prime a^{\prime }})^{2}+1+[2e^{-2\phi }\pi _{x}\pi _{\bar{x}%
}+\pi _{D}^{2}+\sum_{a^{\prime }=5}^{8}(u^{a^{\prime }})^{-2}\pi _{a^{\prime
}}^{2}]  \notag \\
&&+[(\pi _{-}(x^{\prime -}-{\frac{i}{2}(}{\theta }^{i}{{\theta }_{i}^{\prime
}}+{\theta }_{i}{{\theta }^{\prime i})}-{\frac{i}{4}}e^{-2\phi }{(}{\eta }%
^{i}{{\eta }_{i}^{\prime }}+{\eta }_{i}{{\eta }^{\prime i})})+\pi
_{x}x^{\prime }+\pi _{\bar{x}}\bar{x}^{\prime }+\pi _{D}\phi ^{\prime }
\notag \\
&&+\sum_{a^{\prime }=5}^{8}\pi _{a^{\prime }}y^{\prime a^{\prime
}}]^{2}\}+\pi _{-}[{\frac{i}{2}(}{\theta }^{i}{{\dot{\theta}}_{i}}+{\theta }%
_{i}{{\dot{\theta}}^{i})}+{\frac{i}{4}e^{-2\phi }(}{\eta }^{i}{{\dot{\eta}}%
_{i}}+{\eta }_{i}{{\dot{\eta}}^{i})}+{\frac{1}{8}}e^{-2\phi }(\eta ^{i}\eta
_{i})^{2}]  \notag \\
&&-\sum_{a^{\prime }=5}^{8}\pi _{a^{\prime }}{\frac{v^{a^{\prime }}}{{%
u^{a^{\prime }}}}}+v^{9}[\pi _{-}(x^{\prime -}-{\frac{i}{2}(}{\theta }^{i}{{%
\theta }_{i}^{\prime }}+{\theta }_{i}{{\theta }^{\prime i})}-{\frac{i}{4}}%
e^{-2\phi }{(}{\eta }^{i}{{\eta }_{i}^{\prime }}+{\eta }_{i}{{\eta }^{\prime
i})})  \notag \\
&&+\pi _{x}x^{\prime }+\pi _{\bar{x}}\bar{x}^{\prime }+\pi _{D}\phi ^{\prime
}+\sum_{a^{\prime }=5}^{8}\pi _{a^{\prime }}y^{\prime a^{\prime }}].
\label{3-1a}
\end{eqnarray}%
Adding $\pi _{a}$ and$\ \pi _{a^{\prime }}$ as new variables, one can obtain
a new Lagrangian density(Appendix A)
\begin{eqnarray}
\mathcal{\tilde{L}} &=&\tilde{\mathcal{L}}_{k}+\mathcal{L}_{WZ}  \notag \\
&=&\pi _{-}\dot{x}^{-}+\pi _{x}\dot{x}+\pi _{\bar{x}}\dot{\bar{x}}+\pi _{D}%
\dot{\phi}+\sum_{a^{\prime }=5}^{8}\pi _{a^{\prime }}\dot{y}^{a^{\prime }}
\notag \\
&&-\pi _{-}[{\frac{i}{2}(}{\theta }^{i}{{\dot{\theta}}_{i}}+{\theta }_{i}{{%
\dot{\theta}}^{i})}+{\frac{i}{4}e^{-2\phi }(}{\eta }^{i}{{\dot{\eta}}_{i}}+{%
\eta }_{i}{{\dot{\eta}}^{i})]-}\mathcal{H}  \notag \\
&=&f_{i}\dot{z}^{i}+f_{\alpha }\dot{z}^{\alpha }{-}\mathcal{H},  \label{3-2}
\end{eqnarray}%
where $\mathcal{H}$ is the Hamiltonian density and define $H=\int d\sigma
\mathcal{H}$.\ In appendix A\ ,we have proved Lagrangian equations of $%
\mathcal{\tilde{L}}$ include the equations of $\mathcal{L}$ and the
definition of $\pi _{i}.$ We now see that the number of bosonic variables($%
z^{a},\pi _{a},a=1,2,\cdots 8$) and fermionic variables($\theta ^{i},\theta
_{i},\eta ^{i},\eta _{i};i=1,2,3,4.$) are equal. This is an important reason
for requiring the $\kappa $ symmetry.

Therefore all the coefficients $f$ are
\begin{eqnarray}
&&f_{-}=\pi _{-},f_{x}=\pi _{x},f_{\bar{x}}=\pi _{\bar{x}},f_{D}=\pi
_{D},f_{a^{\prime }}=\pi _{a^{\prime }},  \notag \\
&&f_{\theta _{i}}=-{\frac{i}{2}}\pi _{-}\theta ^{i},f_{\theta ^{i}}=-{\frac{i%
}{2}}\pi _{-}\theta _{i},f_{\eta _{i}}=-{\frac{i}{4}}e^{-2\phi }\pi _{-}\eta
^{i},f_{\eta ^{i}}=-{\frac{i}{4}}e^{-2\phi }\pi _{-}\eta _{i},f_{\pi }=0.
\label{3-d}
\end{eqnarray}

\section{Poisson bracket and Jacobi identity}

The Lagrangian(\ref{3-2}) is degenerate in that it is linear in velocities $%
\dot{z}^{a}$. In this section, we study the Lagrangian linear in velocities.
One sees that such system has a natural quasi-symplectic structure, we next
derive the poisson bracket for such degenerate Lagrangian system.

\subsection{Bosonic system}

Assume the\ Lagrangian of bosonic system is

\begin{equation}
L(x_{i},\dot{x}_{i})=\sum\limits_{i}f_{i}(x)\dot{x}_{i}-g(x).
\end{equation}%
The Lagrangian equation is%
\begin{equation*}
\frac{d}{dt}\frac{\partial L}{\partial \dot{x}_{i}}-\frac{\partial L}{%
\partial x_{i}}=0,
\end{equation*}%
giving%
\begin{equation*}
\frac{\partial f_{i}}{\partial x_{j}}\dot{x}_{j}-\frac{\partial f_{j}}{%
\partial x_{i}}\dot{x}_{j}=-\frac{\partial g}{\partial x_{i}}.
\end{equation*}%
Define%
\begin{equation*}
\omega _{ij}=\frac{\partial f_{j}}{\partial x_{i}}-\frac{\partial f_{i}}{%
\partial x_{j}},
\end{equation*}%
we obtain%
\begin{equation}
\partial _{i}\omega _{jk}+cyc(i,j,k)=0,  \label{4-a}
\end{equation}%
and

\begin{equation*}
\omega _{ij}\dot{x}_{j}=\frac{\partial g}{\partial x_{i}}.
\end{equation*}%
If $\omega $ has an inverse $\Omega $
\begin{equation*}
\Omega _{ij}\omega _{jl}=\delta _{il},
\end{equation*}

we have%
\begin{equation*}
\dot{x}_{i}=\Omega _{ij}\frac{\partial g}{\partial x_{j}},
\end{equation*}%
with%
\begin{equation*}
\Omega _{ij}=-\Omega _{ji}.
\end{equation*}
Define poisson bracket
\begin{equation}
\{A,B\}=\frac{\partial A}{\partial x_{i}}\Omega _{ij}\frac{\partial B}{%
\partial x_{j}}.
\end{equation}
Jacobi identity
\begin{equation*}
\{A,\{B,C\}\}+\{B,\{C,A\}\}+\{C,\{A,B\}\}=0,
\end{equation*}%
follows from
\begin{equation*}
\Omega _{il}\partial _{l}\Omega _{jk}+\Omega _{jl}\partial _{l}\Omega
_{ki}+\Omega _{kl}\partial _{l}\Omega _{ij}=0,
\end{equation*}%
due to (\ref{4-a}). One has%
\begin{equation*}
\dot{A}=\{A,g\},
\end{equation*}%
for $A(x).$

\subsection{Bosonic and Fermionic system}

For the system with both bosonic($x_{i}$ Grassmann even) and fermionic($%
\theta _{\alpha }$ Grassmann odd) variables, we have $AB=(-1)^{ab}BA$, $%
\partial _{i}\partial _{j}=(-1)^{ij}\partial _{j}\partial _{i}$, $\partial
_{i}(BC)=(\partial _{i}B)C+(-1)^{ib}B\partial _{i}C.$where $\partial _{i}=%
\frac{\partial }{\partial z_{i}}$, and the Grassmann index of $z_{i}$ is $i$%
, while for $A,B,C$, Grassmann indices are $a,b,c$ respectively (they should
be $\hat{\imath},\hat{\jmath},\hat{a},\hat{b},\hat{c},$ here we abuse the
notation for simplicity).

Assume the Lagrangian is
\begin{equation}
L\mathcal{(}x_{i},\dot{x}_{i},\theta ,\dot{\theta}\mathcal{)=}%
\sum\limits_{i}f_{i}(x,\theta )\dot{x}_{i}+\sum\limits_{\alpha }\psi
_{\alpha }(x,\theta )\dot{\theta}_{\alpha }-g(x,\theta ).
\end{equation}
The Lagrangian equation
\begin{eqnarray*}
\frac{d}{dt}\frac{\partial L}{\partial \dot{x}_{i}}-\frac{\partial L}{%
\partial x_{i}} &=&0, \\
\frac{d}{dt}\frac{\partial L}{\partial \dot{\theta}_{\alpha }}-\frac{%
\partial L}{\partial \theta _{\alpha }}, &=&0,
\end{eqnarray*}
gives
\begin{eqnarray}
\omega _{ij}\dot{x}_{j}+\omega _{i\beta }\dot{\theta}_{\beta } &=&\frac{%
\partial g}{\partial x_{i}},  \notag \\
\omega _{\alpha j}\dot{x}_{j}+\omega _{\alpha \beta }\dot{\theta}_{\beta }
&=&\frac{\partial g}{\partial \theta _{\alpha }},
\end{eqnarray}%
with
\begin{equation*}
\omega _{ij}=\frac{\partial f_{j}}{\partial x_{i}}-\frac{\partial f_{i}}{%
\partial x_{j}}=-\omega _{ji},\omega _{i\beta }=\frac{\partial f_{i}}{%
\partial \theta _{\beta }}+\frac{\partial \psi _{\beta }}{\partial x_{i}}%
=\omega _{\beta j},\omega _{\alpha \beta }=\frac{\partial \psi _{\alpha }}{%
\partial \theta _{\beta }}+\frac{\partial \psi _{\beta }}{\partial \theta
_{\alpha }}=\omega _{\beta \alpha }.
\end{equation*}%
Denote

\begin{equation*}
\omega =\left(
\begin{array}{cc}
\omega _{ij} & \omega _{i\beta } \\
\omega _{\alpha j} & \omega _{\alpha \beta }%
\end{array}%
\right) ,\dot{z}=\left(
\begin{array}{c}
\dot{x}_{j} \\
\dot{\theta}_{\beta }%
\end{array}%
\right) ,\overrightarrow{\partial g}=\left(
\begin{array}{c}
\frac{\partial g}{\partial x_{i}} \\
\frac{\partial g}{\partial \theta _{\alpha }}%
\end{array}%
\right) .
\end{equation*}%
The matrix elements%
\begin{equation*}
\omega _{mn}=\partial _{m}f^{n}-(-1)^{mn+m+n}\partial _{n}f^{m}
\end{equation*}%
satisfies
\begin{equation*}
\partial _{l}\omega _{mn}(-1)^{n^{2}+ln}+cyc(lmn)=0.
\end{equation*}%
If $\omega $ is invertible, one may find $\Omega $, such that $\Omega
_{mn}\omega _{nl}=\delta _{ml}$ giving
\begin{equation*}
\partial _{s}\Omega _{lt}=-\Omega _{lm}\partial _{s}\omega _{mn}\Omega
_{nt}(-1)^{s(l+m)}.
\end{equation*}%
We can further show
\begin{equation*}
\Omega _{mn}=(-1)^{mn+1}\Omega _{mn},
\end{equation*}%
and
\begin{equation*}
\dot{z}_{l}=\Omega _{lm}\partial _{m}g.
\end{equation*}%
In the exponent of (-1),$\ s,m,n,t$ stand for Grassmann indices, they can be
even($i,j$) and odd($\alpha ,\beta $).

Define poisson bracket%
\begin{equation}
\{A,B\}=A\frac{\overleftarrow{\partial }}{\partial z_{l}}\Omega _{lm}\frac{%
\overrightarrow{\partial }B}{\partial z_{m}},
\end{equation}%
one has%
\begin{equation}
\dot{A}=\{A,g\}=A\frac{\overleftarrow{\partial }}{\partial z_{l}}\Omega _{lm}%
\frac{\overrightarrow{\partial }g}{\partial z_{m}}.
\end{equation}

\bigskip Poisson bracket satisfy%
\begin{equation*}
\{A,B\}=(-1)^{ab-1}\{B,A\},
\end{equation*}%
\begin{equation*}
\{A,BC\}=\{A,B\}C+(-1)^{ab}B\{A,C\},
\end{equation*}%
\begin{equation}
\{A,\alpha B+\beta C\}=(-1)^{\alpha a}\alpha \{A,B\}+(-1)^{\beta a}\beta
\{A,C\},\text{ \ for constants }\alpha ,\beta \text{.}
\end{equation}%
here the superscript $a,b,\alpha $ and $\beta $ are Grassmann indices for $%
A,B,\alpha $ and $\beta $ respectively. Super-Jacobi identity is also
satisfied
\begin{equation}
(-1)^{n^{2}+nl}\partial _{l}\omega _{mn}+(-1)^{l^{2}+lm}\partial _{m}\omega
_{nl}+(-1)^{m^{2}+mn}\partial _{n}\omega _{lm}=0,  \label{1}
\end{equation}%
\begin{equation}
\Rightarrow (-1)^{nk}\Omega _{kl}\partial _{l}\Omega _{mn}+(-1)^{km}\Omega
_{ml}\partial _{l}\Omega _{nk}+(-1)^{mn}\Omega _{nl}\partial _{l}\Omega
_{km}=0,  \label{2}
\end{equation}%
\begin{equation}
\Rightarrow
(-1)^{ac}\{A,\{B,C\}\}+(-1)^{ab}\{B,\{C,A\}\}+(-1)^{bc}\{C,\{A,B\}\}=0.
\label{3}
\end{equation}

\subsection{Extension to field theory}

In the following, integration of $\sigma $ over one period is always assumed
if no initial and end points.

The Lagrangian is%
\begin{equation}
L=\int d\sigma f_{i}(z(\sigma ),z^{\prime }(\sigma ))\dot{z}_{i}(\sigma
)-\int d\sigma g(z(\sigma ),z^{\prime }(\sigma )),
\end{equation}
where the index $i$ can be bosonic and fermionic.

The Lagrangian equation%
\begin{equation*}
\frac{d}{dt}(\frac{\delta L}{\delta \dot{z}_{i}(\sigma )})-\frac{\delta L}{%
\delta z_{i}(\sigma )}=0,
\end{equation*}%
gives%
\begin{equation*}
\frac{d}{dt}[(-1)^{j}\int d\sigma ^{\prime }\frac{\delta \dot{z}_{j}(\sigma
^{\prime })}{\delta \dot{z}_{i}(\sigma )}f_{j}(z(\sigma ^{\prime
}),z^{\prime }(\sigma ^{\prime }))]-\int d\sigma ^{\prime }\frac{\delta
f_{j}(z(\sigma ^{\prime }),z^{\prime }(\sigma ^{\prime }))}{\delta
z_{i}(\sigma )}\dot{z}_{j}(\sigma ^{\prime })+\int d\sigma ^{\prime }\frac{%
\delta g(z(\sigma ^{\prime }),z^{\prime }(\sigma ^{\prime }))}{\delta
z_{i}(\sigma )}=0.
\end{equation*}%
Due to%
\begin{equation*}
\frac{\delta \dot{z}_{j}(\sigma ^{\prime })}{\delta \dot{z}_{i}(\sigma )}%
=\delta _{ij}\delta (\sigma ^{\prime }-\sigma ),
\end{equation*}%
one has%
\begin{equation*}
\int d\sigma ^{\prime }\omega _{i(\sigma ),j(\sigma ^{\prime })}\dot{z}%
_{j}(\sigma ^{\prime })=\int d\sigma ^{\prime }\frac{\delta g(z(\sigma
^{\prime }),z^{\prime }(\sigma ^{\prime }))}{\delta z_{i}(\sigma )},
\end{equation*}%
with%
\begin{equation*}
\omega _{i(\sigma ),j(\sigma ^{\prime })}=\frac{\delta f_{j}(z(\sigma
^{\prime }),z^{\prime }(\sigma ^{\prime }))}{\delta z_{i}(\sigma )}%
-(-1)^{i+j(i+j)}\frac{\delta f_{i}(z(\sigma ),z^{\prime }(\sigma ))}{\delta
z_{j}(\sigma ^{\prime })},
\end{equation*}%
which has the property%
\begin{equation*}
\omega _{i(\sigma ),j(\sigma ^{\prime })}=(-1)^{(i-1)(j-1)}\omega _{j(\sigma
^{\prime }),i(\sigma )},
\end{equation*}%
and
\begin{equation}
(-1)^{j^{2}+jl}\frac{\delta }{\delta z_{l}(\sigma _{l})}\omega _{i(\sigma
_{i}),j(\sigma _{j})}+(-1)^{l^{2}+il}\frac{\delta }{\delta z_{i}(\sigma _{i})%
}\omega _{j(\sigma _{j}),l(\sigma _{l})}+(-1)^{i^{2}+ij}\frac{\delta }{%
\delta z_{j}(\sigma _{j})}\omega _{l(\sigma _{l}),i(\sigma _{i})}=0.
\label{1+}
\end{equation}

Assume matrix [$\omega _{i(\sigma ),j(\sigma ^{\prime })}$] has an inverse [$%
\ \Omega _{j(\sigma ^{\prime }),i(\sigma )}$]%
\begin{eqnarray*}
\int d\sigma ^{\prime }\omega _{i(\sigma ),j(\sigma ^{\prime })}\Omega
_{j(\sigma ^{\prime }),l(\sigma ^{\prime \prime })} &=&\delta _{il}\delta
(\sigma -\sigma ^{\prime \prime }), \\
\int d\sigma \Omega _{i(\sigma ^{\prime }),j(\sigma )}\omega _{j(\sigma
),l(\sigma ^{^{\prime \prime }})} &=&\delta _{il}\delta (\sigma ^{\prime
}-\sigma ^{\prime \prime }),
\end{eqnarray*}%
Then we have%
\begin{equation*}
\Omega _{i(\sigma ),j(\sigma ^{\prime })}=(-1)^{ij+1}\Omega _{j(\sigma
^{\prime }),i(\sigma )}.
\end{equation*}

The Lagrangian equation
\begin{equation}
\int d\sigma ^{\prime }\omega _{i(\sigma ),j(\sigma ^{\prime })}\dot{z}%
_{j}(\sigma ^{\prime })=\int d\sigma ^{\prime }\frac{\delta g(z(\sigma
^{\prime }),z^{\prime }(\sigma ^{\prime }))}{\delta z_{i}(\sigma )}=\frac{%
\delta H}{\delta z_{i}(\sigma )},  \label{5-1}
\end{equation}%
gives%
\begin{equation*}
\dot{z}_{k}(\sigma ^{\prime \prime })=\int d\sigma \Omega _{k(\sigma
^{\prime \prime }),i(\sigma )}\frac{\delta H}{\delta z_{i}(\sigma )},
\end{equation*}%
where $H=\int d\sigma g(z(\sigma ),z^{\prime }(\sigma )).$

Define Poisson bracket
\begin{eqnarray}
\{A,B\} &=&\int d\sigma \int d\sigma ^{\prime }A\frac{\overleftarrow{\delta }%
}{\delta z_{i}(\sigma )}\Omega _{i(\sigma ),j(\sigma ^{\prime })}\frac{%
\overrightarrow{\delta }B}{\delta z_{j}(\sigma ^{\prime })}  \notag \\
&=&\int d\sigma \int d\sigma ^{\prime }(-1)^{(a-1)i}\frac{\delta A}{\delta
z_{i}(\sigma )}\Omega _{i(\sigma ),j(\sigma ^{\prime })}\frac{\delta B}{%
\delta z_{j}(\sigma ^{\prime })},
\end{eqnarray}%
which implies
\begin{equation*}
\{A,B\}=(-1)^{ab+1}\{B,A\}.
\end{equation*}

We have
\begin{equation*}
\dot{z}_{k}(\sigma ^{\prime \prime })=\{z_{k}(\sigma ^{\prime \prime
}),H\}=\int d\sigma \int d\sigma ^{\prime }z_{k}(\sigma ^{\prime \prime })%
\frac{\overleftarrow{\delta }}{\delta z_{i}(\sigma )}\Omega _{i(\sigma
),j(\sigma ^{\prime })}\frac{\delta H}{\delta z_{j}(\sigma ^{\prime })},
\end{equation*}%
and
\begin{equation}
\dot{A}=\{A,H\}.  \label{35}
\end{equation}

The l.h.s. of Lagrangian equation (\ref{5-1})\ can be written as

\begin{eqnarray*}
\int d\sigma \omega _{i(\sigma ),j(\sigma ^{\prime })}\dot{z}_{j}(\sigma
^{\prime }) &=&\int d\sigma ^{\prime }\{\delta (\sigma ^{\prime }-\sigma )%
\frac{\partial f_{j}}{\partial z_{i}}(\sigma ^{\prime })-(-1)^{i+j+ij}\delta
(\sigma -\sigma ^{\prime })\frac{\partial f_{i}}{\partial z_{j}}(\sigma ) \\
&&+{\frac{\partial }{\partial _{\sigma ^{\prime }}}}\delta (\sigma ^{\prime
}-\sigma ){\frac{\partial f_{j}}{\partial z_{i}^{\prime }}}(\sigma ^{\prime
})-(-1)^{i+j+ij}{\frac{\partial }{\partial _{\sigma }}}\delta (\sigma
-\sigma ^{\prime }){\frac{\partial f_{i}}{\partial z_{j}^{\prime }}}(\sigma
)\}\dot{z}_{j}(\sigma ^{\prime }).
\end{eqnarray*}%
Using integration by parts and considering $\delta (\sigma -\sigma ^{\prime
})=\delta (\sigma ^{\prime }-\sigma )$ and ${\frac{\partial }{\partial
_{\sigma ^{\prime }}}}\delta (\sigma ^{\prime }-\sigma )=-{\frac{\partial }{%
\partial _{\sigma }}}\delta (\sigma -\sigma ^{\prime }),$\ we have
\begin{equation*}
\int d\sigma ^{\prime }\omega _{i(\sigma ),j(\sigma ^{\prime })}\dot{z}%
_{j}(\sigma ^{\prime })=\tilde{\omega}_{ij(\sigma )}\dot{z}_{j}(\sigma )+%
\mathcal{A}_{ij}\dot{z}_{j}^{\prime }(\sigma ),
\end{equation*}%
where
\begin{eqnarray*}
\tilde{\omega}_{ij(\sigma )} &=&{\frac{\partial f_{j}}{\partial z_{i}}}%
(\sigma )-(-1)^{i+j+ij}{\frac{\partial f_{i}}{\partial z_{j}}}(\sigma )-[{%
\frac{\partial f_{j}}{\partial z_{i}^{\prime }}}(\sigma )]^{\prime } \\
\mathcal{A}_{ij}(\sigma ) &=&-[{\frac{\partial f_{j}}{\partial z_{i}^{\prime
}}}(\sigma )+(-1)^{i+j+ij}{\frac{\partial f_{i}}{\partial z_{j}^{\prime }}}%
(\sigma )].
\end{eqnarray*}%
Locality condition requires $\mathcal{A}_{ij}=0$. If this condition is
satisfied, we may instead use
\begin{equation*}
\omega _{i(\sigma ),j(\sigma ^{\prime })}=\tilde{\omega}_{ij(\sigma )}\delta
(\sigma -\sigma ^{\prime }),
\end{equation*}%
and
\begin{equation*}
\Omega _{i(\sigma ),j(\sigma ^{\prime })}=\tilde{\Omega}_{ij(\sigma )}\delta
(\sigma -\sigma ^{\prime }).
\end{equation*}%
for the inverse of $\omega _{i(\sigma ),j(\sigma ^{\prime })}$ to get the
correct equation of motion. We may also define poisson bracket as%
\begin{equation}
\{A,B\}=\int d\sigma d\sigma ^{\prime }[A\frac{\overleftarrow{\delta }}{%
\delta z_{i}(\sigma )}\tilde{\Omega}_{ij}(\sigma )\delta (\sigma -\sigma
^{\prime })\frac{\overrightarrow{\delta }B}{\delta z_{j}(\sigma ^{\prime })}%
],
\end{equation}%
and show that the Jacobi identity is still valid for the case we are
interested in in this paper(Appendix B).

On the contrary, for $\mathcal{A}_{ij}\neq 0$, we find the locality of the
field theory is broken. This is because the equation gives
\begin{equation}
\dot{z}_{i}^{\prime }=M_{ij}\dot{z}_{j}+N_{i}.
\end{equation}%
Even though $M_{ij}$ and $N_{i}$ are local at each $\sigma $. The quantity $%
\dot{z}_{j}(\sigma )$ is determined by all data of $M_{ij}$ and $N_{i}$ at $%
\sigma ^{\prime }=\sigma _{0}$ to $\sigma $, if $\dot{z}_{j}(\sigma _{0})$
is given at point $\sigma _{0}$. Thus the 2-D field theory is nonlocal.

For the system described by (\ref{3-2}), we see that from (\ref{3-d})
\begin{equation}
{\frac{\partial f_{a}}{\partial z^{\prime b}}}=0,
\end{equation}%
for $a=i,\alpha $, thus it does satisfies
\begin{equation}
\mathcal{A}_{ab}\equiv -{\frac{\partial f_{b}}{\partial z^{\prime a}}}%
-(-1)^{a+b+ab}{\frac{\partial f_{a}}{\partial z^{\prime b}}}=0.
\end{equation}%
The problem left is whether the supermatrix $\tilde{\Omega}_{ij(\sigma )}$
is invertible. Further calculation of $\omega _{ab}$ via
\begin{equation}
\tilde{\omega}_{ab}=(-1)^{a+b+ab}{\partial _{b}f_{a}}+\partial _{a}f_{b},
\end{equation}%
for (\ref{3-2}) gives:

(A) $\tilde{\omega}_{ij}=-\tilde{\omega}_{ji}$
\begin{equation}
\tilde{\omega}_{x^{-}\pi _{-}}=-1,\tilde{\omega}_{x\pi _{x}}=-1,\tilde{\omega%
}_{D\pi _{D}}=-1,\tilde{\omega}_{a^{\prime }\pi _{a^{\prime }}}=-1.
\end{equation}

(B) $\tilde{\omega}_{i\alpha }=\tilde{\omega}_{\alpha i}$
\begin{eqnarray}
\tilde{\omega}_{\pi _{-}\theta _{i}} &=&-{\frac{i}{2}}\theta ^{i},\tilde{%
\omega}_{\pi _{-}\theta ^{i}}=-{\frac{i}{2}}\theta _{i},\tilde{\omega}_{\pi
_{-}\eta _{i}}=-{\frac{i}{4}}e^{-2\phi }\eta ^{i}, \\
\tilde{\omega}_{\pi _{-}\eta _{i}} &=&-{\frac{i}{4}}e^{-2\phi }\eta ^{i},%
\tilde{\omega}_{\phi \eta _{i}}={\frac{i}{2}}e^{-2\phi }\pi _{-}\eta ^{i},%
\tilde{\omega}_{\phi \eta ^{i}}={\frac{i}{2}}e^{-2\phi }\pi _{-}\eta _{i}.
\end{eqnarray}

(C) $\tilde{\omega}_{\alpha \beta }=\tilde{\omega}_{\beta \alpha }$
\begin{equation}
\tilde{\omega}_{\theta _{i}\theta ^{i}}=-{\frac{i}{2}}\pi _{-},\tilde{\omega}%
_{\theta ^{i}\theta _{i}}=-{\frac{i}{2}}\pi _{-},\tilde{\omega}_{\eta
_{i}\eta ^{i}}=-{\frac{i}{4}}e^{-2\phi }\pi _{-},\tilde{\omega}_{\eta
^{i}\eta _{i}}=-{\frac{i}{4}}e^{-2\phi }\pi _{-}.
\end{equation}

We conclude that as long as the c number part of $\pi _{-}\neq 0$, $\omega $
is invertible, and the Poisson bracket is well defined. On the other hand,
we have from Appendix A,
\begin{equation*}
\pi _{-}=e^{\phi }J_{-}=e^{\phi }{\frac{\partial \mathcal{L}_{k}}{\partial
x_{0}^{-}}}={\frac{e^{2\phi }}{\sqrt{-\mathcal{G}}}}[e^{2\phi }((x^{1\prime
})^{2}+(x^{2\prime })^{2})+{\phi ^{\prime }}^{2}+\sum_{a^{\prime
}=5}^{8}(u^{a^{\prime }}y^{\prime {a^{\prime }}})^{2}+1],
\end{equation*}%
and
\begin{equation}
{\frac{1}{\pi _{-}}}={\frac{{e^{-2\phi }\sqrt{-\mathcal{G}}}}{e^{2\phi
}(x_{1}^{\prime 2}+x_{2}^{\prime 2})+{\phi ^{\prime }}^{2}+\sum_{a^{\prime
}=5}^{8}(u^{a^{\prime }}y^{\prime {a^{\prime }}})^{2}+1}.}
\end{equation}%
Thus as long as $\mathcal{G}$ exits, we have a well defined Poisson bracket.
This is a loose condition. This condition may break down, for example, when
the "string tube" grows a new branch in the $AdS_{5}\times S_{5}$ space. But
in most cases, the Hamiltonian description is valid. The difficulty may
appear in the quantum theory, where one has to take into account the whole
space time. This needs further investigation.

\section{Flat currents and solution transformation}

\subsection{Flat currents with one parameter}

Bena, Polchinski and Roiban made an important discovery that the Metsaev and
Tseytlin model has a one-parameter family of flat currents. This implies the
model has infinite conserved nonlocal quantities. Here we review the
equivalent form of their construction.

From $\mathcal{J}=G^{\prime -1}dG^{\prime }=L^{a}P_{a}+L^{a^{\prime
}}P_{a^{\prime }}\,+\frac{1}{2}L^{ab}J_{ab}\,+\frac{1}{2}L^{a^{\prime
}b^{\prime }}J_{a^{\prime }b^{\prime }}\,+L^{\alpha \alpha ^{\prime
}I}Q_{\alpha \alpha ^{\prime }}^{I},$one has $d\mathcal{J}+\mathcal{J}\wedge
\mathcal{J}=0,$ giving the Maurer-Cartan equations

\begin{align}
dL^{a}& =-L^{b}\wedge L^{ba}-\mathrm{i}\bar{L}^{I}\gamma ^{a}\wedge L^{I},
\notag \\
dL^{a^{\prime }}& =-L^{b^{\prime }}\wedge L^{b^{\prime }a^{\prime }}+\bar{L}%
^{I}\gamma ^{a^{\prime }}\wedge L^{I},  \notag \\
dL^{ab}& =-L^{a}\wedge L^{b}-L^{ac}\wedge L^{cb}+\epsilon _{IJ}\bar{L}%
^{I}\gamma ^{ab}\wedge L^{J},  \notag \\
dL^{a^{\prime }b^{\prime }}& =L^{a^{\prime }}\wedge L^{b^{\prime
}}-L^{a^{\prime }c^{\prime }}\wedge L^{c^{\prime }b^{\prime }}-\epsilon _{IJ}%
\bar{L}^{I}\gamma ^{a^{\prime }b^{\prime }}\wedge L^{J},  \notag \\
dL^{I}& =-\frac{\mathrm{i}}{2}\gamma _{a}\epsilon _{IJ}L^{J}\wedge L^{a}+%
\frac{1}{2}\gamma _{a^{\prime }}\epsilon _{IJ}L^{J}\wedge L^{a^{\prime }}
\notag \\
& +\frac{1}{4}\gamma _{ab}L^{I}\wedge L^{ab}+\frac{1}{4}\gamma _{a^{\prime
}b^{\prime }}L^{I}\wedge L^{a^{\prime }b^{\prime }}.  \label{mc}
\end{align}

We firstly introduce the world-sheet Hodge dual of the Maurer-Cartan 1-forms
$L^{a}$ and $L^{a^{\prime }}$. Let $\sqrt{-g}g^{ij}=\gamma ^{ij}$ and

\begin{equation*}
^{\ast }L_{k}^{\hat{a}}=-\epsilon ^{ki}\gamma ^{ij}L_{j}^{\hat{a}},\text{and
}\hat{a}=a,a^{\prime },\text{\ }\epsilon ^{01}=-\epsilon ^{10}=1.\text{\ }
\end{equation*}%
The equations of motion can be expressed as
\begin{eqnarray}
d^{\mathbf{\ast }}L^{a}+L^{ab}\wedge ^{\ast }L^{b}+\mathrm{i}s^{IJ}\bar{L}%
^{I}\gamma ^{a}\wedge L^{J} &=&0,  \label{EM1} \\
d^{\mathbf{\ast }}L^{a^{\prime }}+L^{a^{\prime }b^{\prime }}\wedge ^{\ast
}L^{b^{\prime }}-s^{IJ}\bar{L}^{I}\gamma ^{a^{\prime }}\wedge L^{J} &=&0,
\label{EM2} \\
\delta ^{IJ}(^{\ast }L^{a}\gamma ^{a}+\mathrm{i}^{\ast }L^{a^{\prime
}}\gamma ^{a^{\prime }})\wedge L^{J}+s^{IJ}(L^{a}\gamma ^{a}+\mathrm{i}%
L^{a^{\prime }}\gamma ^{a^{\prime }})\wedge L^{J} &=&0.  \label{EM 3}
\end{eqnarray}

Introduce the forms with a parameter $\lambda $ ,
\begin{eqnarray}
L^{a}\left( \lambda \right) &=&\frac{1}{2}\left( \lambda ^{2}+\lambda
^{-2}\right) L^{a}+\frac{1}{2}\left( \lambda ^{2}-\lambda ^{-2}\right)
\,^{\ast }L^{a},  \notag \\
L^{a^{\prime }}\left( \lambda \right) &=&\frac{1}{2}\left( \lambda
^{2}+\lambda ^{-2}\right) L^{a^{\prime }}+\frac{1}{2}\left( \lambda
^{2}-\lambda ^{-2}\right) \,^{\ast }L^{a^{\prime }},  \notag \\
L^{ab}\left( \lambda \right) &=&L^{ab},\quad L^{a^{\prime }b^{\prime
}}\left( \lambda \right) =L^{a^{\prime }b^{\prime }},  \notag \\
L^{1}\left( \lambda \right) &=&\lambda L^{1},\quad L^{2}\left( \lambda
\right) =\lambda ^{-1}L^{2}.  \label{td}
\end{eqnarray}%
When $\det[\gamma _{ij}]=-1,$ we have
\begin{eqnarray}
^{\ast }(^{\ast }A) &=&A,  \notag \\
A\wedge \,^{\ast }B &=&-\,^{\ast }A\wedge B,  \notag \\
^{\ast }A\wedge \,^{\ast }B &=&-\,A\wedge B.  \label{5-2}
\end{eqnarray}%
We can prove one forms (\ref{td}) with a parameter also satisfy
Maurer-Cartan equations(\ref{mc}) by (\ref{EM1}) to (\ref{5-2}). Thus the
currents $\mathcal{J}(\lambda )$ with spectral parameter $\lambda $
expressed from the Cartan one forms(\ref{td}),
\begin{equation}
\mathcal{J}(\lambda )=L^{a}(\lambda )P_{a}+L^{a^{\prime }}(\lambda
)P_{a^{\prime }}\,+\frac{1}{2}L^{ab}(\lambda )J_{ab}\,+\frac{1}{2}%
L^{a^{\prime }b^{\prime }}(\lambda )J_{a^{\prime }b^{\prime }}\,+L^{\alpha
\alpha ^{\prime }I}(\lambda )Q_{\alpha \alpha ^{\prime }}^{I},  \label{5-4}
\end{equation}
satisfies $d\mathcal{J}(\lambda )+\mathcal{J}(\lambda )\wedge \mathcal{J}%
(\lambda )=0$. So the equation $\mathcal{J}(\lambda )=G(\lambda
)^{-1}dG(\lambda )$ is integrable and $\mathcal{J}(\lambda )$ naturally
leads to an infinite number of non-local conserved quantities.

From the above, we see that as long as equations of motion are satisfied, $%
\mathcal{J}(\lambda )$ will be flat. However, after some gauge fixing, some
equations of motion are missing. The $\kappa $ symmetry gauge fixing cause
half of equations(\ref{EM 3}) disappear and fixing $x_{0}=\tau ,y_{9}=\sigma
$ cause two of (\ref{EM1},\ref{EM2}) missing. Are they still satisfied? The
answer is affirmative. The reason is that local symmetry cause $\delta S=0,$
under certain combination of canonical variables. Thus equations of motion
are not independent. The number of redundancy of them exactly matches the
number of missing equations in the gauge fixing. In references \cite{AF2},
the authors give a concise proof that the flat currents keeps flat under
various symmetry transformations. This explains the origin of that\ the flat
currents still exists after gauge fixing.

We then express the flat currents in terms of canonical variables for the
system in section 3. One has
\begin{equation*}
^{\ast }L_{k}^{\hat{a}}=-\epsilon ^{ki}\gamma ^{ij}L_{j}^{\hat{a}}.
\end{equation*}%
From (\ref{in}) and (\ref{3-f}), we have
\begin{eqnarray*}
\delta S_{k} &=&-{\frac{1}{2}}\int d\tau d\sigma \delta \gamma ^{\mu \nu
}x_{\mu }^{\hat{a}}G_{\hat{a}\hat{b}}x_{\nu }^{\hat{b}}-{\frac{1}{2}}\int
d\tau d\sigma \gamma ^{\mu \nu }\delta x_{\mu }^{\hat{a}}G_{\hat{a}\hat{b}%
}x_{\nu }^{\hat{b}}-{\frac{1}{2}}\int d\tau d\sigma \gamma ^{\mu \nu }x_{\mu
}^{\hat{a}}G_{\hat{a}\hat{b}}\delta x_{\nu }^{\hat{b}} \\
&=&-\int d\tau d\sigma \gamma ^{\mu \nu }\delta x_{\mu }^{\hat{a}}G_{\hat{a}%
\hat{b}}x_{\nu }^{\hat{b}},
\end{eqnarray*}%
giving
\begin{equation*}
{\frac{\partial \mathcal{L}_{k}}{\partial x_{0}^{\hat{a}}}}=-\gamma ^{0\nu
}G_{\hat{a}\hat{b}}x_{\nu }^{\hat{b}}=J_{\hat{a}},
\end{equation*}%
and%
\begin{equation*}
^{\ast }L_{1}^{\hat{a}}=\epsilon ^{10}(G^{-1})^{\hat{a}\hat{b}}J_{\hat{a}%
}=-(G^{-1})^{\hat{a}\hat{b}}J_{\hat{a}}.
\end{equation*}%
From appendix A, $J_{\hat{a}}$\ is expressed in ($\pi _{i},z^{\prime
i},z^{i},z^{\prime \alpha },z^{\alpha }$) for all ten $\hat{a}$'s$.$ Further
checking the remaining components of $\mathcal{J}_{1}(\lambda )$ by (\ref%
{8-20}), we find $\mathcal{J}_{1}(\lambda )$ can be expressed by these
variables too.

\subsection{solution transformations}

For the flat currents
\begin{equation}
\mathcal{J}(\lambda )=L^{a}(\lambda )P_{a}+L^{a^{\prime }}(\lambda
)P_{a^{\prime }}\,+\frac{1}{2}L^{ab}(\lambda )J_{ab}\,+\frac{1}{2}%
L^{a^{\prime }b^{\prime }}(\lambda )J_{a^{\prime }b^{\prime }}\,+L^{\alpha
\alpha ^{\prime }I}(\lambda )Q_{\alpha \alpha ^{\prime }}^{I},  \label{6-1}
\end{equation}%
the solution $G(\lambda ,\tau ,\sigma )^{-1}\partial _{\mu }G(\lambda ,\tau
,\sigma )=\mathcal{J}_{\mu }(\lambda )$ $(\mu =0,1)$ with given $G(\lambda
,\tau _{0},\sigma _{0})=G_{0}$ is independent of the path of integration. It
can be symbolically expressed as
\begin{equation*}
G_{0}\mathrm{P}e^{\int _{C}\mathcal{J}(\lambda )}=G(\lambda ,\tau ,\sigma ),
\end{equation*}%
where $C$ is any contour from $(\tau _{0},\sigma _{0})$ to $(\tau ,\sigma )$%
, and $\mathrm{P}$ denotes path ordering of the Lie algebra generators.
Consider two paths ABC and ADC, where AB and DC are along $\sigma$ with
length $L$ while BC and AD are along $\tau$. We have $G(\lambda ,\tau
_{c},\sigma _{c})=G_{0}U_{AB}U_{BC}=G_{0}U_{AD}U_{DC}$, where
\begin{eqnarray*}
U_{AB} &=&\mathrm{P}e_{A}^{\int^{B}\mathcal{J}(\lambda )},U_{BC}=\mathrm{P}%
e_{B}^{\int^{C}\mathcal{J}(\lambda )}, \\
U_{AD} &=&\mathrm{P}e_{A}^{\int^{D}\mathcal{J}(\lambda )},U_{DC}=\mathrm{P}%
e_{D}^{\int^{C}\mathcal{J}(\lambda )}.
\end{eqnarray*}
If the period of $\sigma $ is $L$, $U_{AD}$ and $U_{BC}$ are equal. We have%
\begin{equation*}
U_{DC}=U_{AD}^{-1}U_{AB}U_{BC}=U_{BC}^{-1}U_{AB}U_{BC}.
\end{equation*}
Thus in any representation of $PSU(2,2|4)$, the matrices $\hat{U}_{DC}$ and\
$\hat{U}_{AB}$ are similar matrices and the supertraces of them are equal.
That is, $F_{1}(\lambda )=\mathbf{str}\mathrm{P}e_{0}^{\int^{L} \hat{%
\mathcal{J}}_{1}(\lambda ,\sigma ,\tau )}=\mathbf{str}\hat{U}(\tau )$ is a
constant of motion\cite{BPR03} as well as their eigenvalues\cite{AFT}. Let's
return to the system(\ref{3-2}) in section 3,\ since $\mathcal{J}
_{1}(\lambda )$ is a function of canonical variables and $F_{1}(\lambda )$\
is always conserved, we have
\begin{equation}
\{F_{1}(\lambda ),H\}=0
\end{equation}%
by (\ref{35}). Notice $F_{1}(\lambda )$ and$\ H$ are not depending on $\tau
=x^{+}$ when expressed by $\pi _{i},z_{i},z_{i}^{\prime },z_{\alpha
},z_{\alpha }^{\prime }$ in section 3.

Due to Jacobi identity, we have
\begin{equation*}
\{z_{i},\{F_{1}(\lambda ),H\}\}+\{F_{1}(\lambda
),\{H,z_{i}\}\}+\{H,\{z_{i},F_{1}(\lambda )\}\}=0,
\end{equation*}%
\begin{equation*}
\{\{z_{i},H\},F_{1}(\lambda )\}=\{\{z_{i},F_{1}(\lambda )\},H\},
\end{equation*}%
impling\ the action of Hamiltonian $H$ and the action of $F_{1}(\lambda )$
are commutable\cite{Arnold}. Assume a solution $z(\tau ,\sigma )$ is given
by $z(0,\sigma )$ and satisfies
\begin{equation}
\dot{z}_{i}=\{z_{i},H\},  \label{solu}
\end{equation}%
where $z_{i}$ can be bosonic and fermionic variables.

We may solve%
\begin{equation}
\frac{d}{dt}z_{i}=\{z_{i},F_{1}(\lambda )\},
\end{equation}
with $z_{i}(\lambda ,t=0,\tau ,\sigma )=z_{i}(\tau ,\sigma ).$ Then $%
z_{i}(\lambda ,t,\tau ,\sigma )$ is a new solution of (\ref{solu}) for each
fixed $t$. This is a solution transformation. There are infinite generators
of such transformations.

\section{Discussions}

In this paper, we construct the solution transformations by Jacobi identity
of poisson bracket for one parameter flat currents with Hamiltonian, and the
poisson bracket is constructed from Lagrangian which is linear in velocities
with $\kappa$ Light-cone gauge fixing. The relation of solution
transformations for different $\lambda ,t$ is not clear, it seems that they
form two parameter sets. Since the expression of $F_{1}$ is complicated, the
further investigation of examples is worth doing.

\section*{Acknowledgements}

We are grateful to professor B.Y. Hou and professor R. H. Yue for helpful
discussions and suggestions. This work is supported by National Natural
Science foundation of China under Grant No. 10575080. Z. Y. Wang and J. Feng
are also supported by the NWU Graduate Cross-discipline Funds(08YJC24).

\appendix

\section{Hamiltonian Analysis}

Let $\mathcal{L}=\mathcal{L}_{k}+\mathcal{L}_{WZ},$ $z=z_{i},z_{\alpha }$
and c number of $\det (\frac{\partial ^{2}\mathcal{L}_{k}}{{\partial \dot{z}%
_{i}\partial \dot{z}_{j}}})\neq 0,$define $\pi _{i}={\frac{\partial {%
\mathcal{L}_{k}}}{\partial \dot{z}_{i}},}$ $\mathcal{\tilde{H}}=\pi _{i}\dot{%
z}_{i}-\mathcal{L}_{k}.$ One can express $\dot{z}_{i}$ and $\mathcal{\tilde{H%
}}$ as the functions of$\ \pi _{i},z_{i},z_{i}^{\prime },z_{\alpha
},z_{\alpha }^{\prime },\dot{z}_{\alpha }.$ We have a new Lagrangian density
\begin{equation*}
\mathcal{\tilde{L}}_{k}=\pi _{i}\dot{z}_{i}-\mathcal{\tilde{H}}.
\end{equation*}

The variation of $\mathcal{\tilde{H}}$ is
\begin{equation}
\delta \mathcal{\tilde{H}}=\delta \pi _{i}\dot{z}_{i}+\pi _{i}\delta \dot{z}%
_{i}-{\frac{\partial \mathcal{L}_{k}}{\partial \dot{z}_{i}}}\delta \dot{z}%
_{i}-{\frac{\partial \mathcal{L}_{k}}{\partial z_{i}}}\delta z_{i}-{\frac{%
\partial \mathcal{L}_{k}}{\partial z_{i}^{\prime }}}\delta z_{i}^{\prime }-{%
\frac{\partial \mathcal{L}_{k}}{\partial z_{\alpha }}}\delta z_{\alpha }-{%
\frac{\partial \mathcal{L}_{k}}{\partial \dot{z}_{\alpha }}}\delta \dot{z}%
_{\alpha }-{\frac{\partial \mathcal{L}_{k}}{\partial z_{\alpha }^{\prime }}}%
\delta z_{\alpha }^{\prime }.  \label{A1}
\end{equation}%
Denote
\begin{equation*}
L_{u}f\equiv {\frac{\partial f}{\partial z_{i}}}-{\frac{\partial }{\partial
\tau }}{\frac{\partial f}{\partial \dot{u}}}-{\frac{\partial }{\partial
\sigma }}{\frac{\partial f}{\partial u^{\prime }}.}
\end{equation*}

The equation $L_{\pi _{i}}\mathcal{\tilde{L}}_{k}=0$ gives
\begin{equation*}
(\pi _{j}-{\frac{\partial {\mathcal{L}_{k}}}{\partial \dot{z}_{j}}}){\frac{%
\partial \dot{z_{j}}}{\partial \pi _{i}}}=0,
\end{equation*}
implying
\begin{equation}
\pi _{j}-{\frac{\partial {\mathcal{L}_{k}}}{\partial \dot{z}_{j}}}=0,
\label{A2}
\end{equation}%
when c number of $\det (\frac{\partial ^{2}\mathcal{L}_{k}}{{\partial \dot{z}%
_{i}\partial \dot{z}_{j}}})\neq 0$. We have
\begin{equation*}
L_{z_{i}}\tilde{L}_{k}\equiv {\frac{\partial \mathcal{\tilde{L}}_{k}}{%
\partial z_{i}}}-{\frac{\partial }{\partial \tau }}{\frac{\partial \mathcal{%
\tilde{L}}_{k}}{\partial \dot{z}_{i}}}-{\frac{\partial }{\partial \sigma }}{%
\frac{\partial \mathcal{\tilde{L}}_{k}}{\partial z_{i}^{\prime }}}=-{\frac{%
\partial \mathcal{\tilde{H}}}{\partial z_{i}}}-{\frac{\partial \pi _{i}}{%
\partial \tau }}+{\frac{\partial }{\partial \sigma }}{\frac{\partial
\mathcal{\tilde{H}}}{\partial z_{i}^{\prime }}}=L_{z_{i}}{}\mathcal{L}_{k},
\end{equation*}

\begin{equation*}
L_{z_{\alpha }}\tilde{L}_{k}\equiv {\frac{\partial \mathcal{\tilde{L}}_{k}}{%
\partial z_{\alpha }}}-{\frac{\partial }{\partial \tau }}{\frac{\partial
\mathcal{\tilde{L}}_{k}}{\partial \dot{z}_{\alpha }}}-{\frac{\partial }{%
\partial \sigma }}{\frac{\partial \mathcal{\tilde{L}}_{k}}{\partial
z_{\alpha }^{\prime }}}=-{\frac{\partial \mathcal{\tilde{H}}}{\partial
z_{\alpha }}}+{\frac{\partial }{\partial \tau }(\frac{\partial \mathcal{%
\tilde{H}}}{\partial \dot{z}_{\alpha }})}+{\frac{\partial }{\partial \sigma }%
}{\frac{\partial \mathcal{\tilde{H}}}{\partial z_{\alpha }^{\prime }}}%
=L_{z_{\alpha }}\mathcal{L}_{k}.
\end{equation*}%
The last step comes from (\ref{A1}) and (\ref{A2}).

For $\mathcal{L}=\mathcal{L}_{k}+\mathcal{L}_{WZ}$ and $\tilde{\mathcal{L}}=%
\mathcal{\tilde{L}}_{k}+\mathcal{L}_{WZ}$, we have
\begin{eqnarray}
0 &=&L_{\pi _{i}}\mathcal{\tilde{L}}=L_{\pi _{i}}\mathcal{\tilde{L}}%
_{k}\Rightarrow \pi _{i}-{\frac{\partial {\mathcal{L}_{k}}}{\partial \dot{z}%
_{i}}}=0,  \notag \\
0 &=&L_{z_{i}}\mathcal{\tilde{L}}=L_{z_{i}}\mathcal{\tilde{L}}_{k}+L_{z_{i}}%
\mathcal{L}_{WZ}=L_{z_{i}}\mathcal{L}_{k}+L_{z_{i}}\mathcal{L}%
_{WZ}\Rightarrow L_{z_{i}}{}\mathcal{L}=0,  \notag \\
0 &=&L_{z_{\alpha }}\mathcal{\tilde{L}}=L_{z_{\alpha }}\mathcal{\tilde{L}}%
_{k}+L_{z_{\alpha }}\mathcal{L}_{WZ}=L_{z_{\alpha }}\mathcal{L}%
_{k}+L_{z_{\alpha }}\mathcal{L}_{WZ}\Rightarrow L_{z_{\alpha }}{}\mathcal{L}%
=0.
\end{eqnarray}%
Thus the Lagrangian equations of $\tilde{\mathcal{L}}$ with the variables ($%
\pi _{i},\dot{z}_{i},z_{i}^{\prime },z_{i},\dot{z}_{\alpha },z_{\alpha
}^{\prime },z_{\alpha }$) give the definition of $\pi _{i}$ and the
Lagrangian equation of $\mathcal{L}$ for variables ($\dot{z}%
_{i},z_{i}^{\prime },z_{i},\dot{z}_{\alpha },z_{\alpha }^{\prime },z_{\alpha
}$).

This is a partial Legendre transformation for the part Lagrangian with
arbitrary partition of $\mathcal{L}=\mathcal{L}_{1}+\mathcal{L}_{2}$ . This
idea is firstly introduced by Arutyunov, Frolov et al.{\cite{AF2} in
deriving the Hamiltonian of GS string. }

For the Polyakov Lagrangian(\ref{3-f}), the variation of $g^{\mu \nu }$
yields the well known Nambu-Goto Lagrangian,
\begin{equation*}
\mathcal{L}_{k}=-{\sqrt{%
(x_{0}^{a}G_{ab}x_{1}^{b})^{2}-(x_{0}^{a}G_{ab}x_{0}^{b})(x_{1}^{a}G_{ab}x_{1}^{b})%
}\equiv -\sqrt{-\mathcal{G}}},
\end{equation*}%
where $a,b=+,-,x,\bar{x},D$ and $A^{\prime }(A^{\prime }=5,6,7,8,9).$ Then
we fix the gauge $x^{+}=\tau $, $y^{9}=\sigma $, and write
\begin{equation*}
L_{\mu }^{a}\equiv x_{\mu }^{a}=x_{\mu 0}+\alpha _{i}^{a}z_{\mu }^{i},
\end{equation*}%
where $z_{\mu }^{i}=\label{eq:1}\left\{
\begin{aligned}
         \dot z^i, \mu=0 \\
             z^{\prime i}, \mu=1
                          \end{aligned}\right. $, $i\neq x^{+},y^{9}$, and $%
x_{\mu 0}^{a}$ is the rest of the $L_{\mu }^{a}$, including fermionic
variables and some functions of coordinates.

We have for $z^{i}=x^{-},x,\bar{x},D,y^{5},y^{6},y^{7},y^{8}$,
\begin{eqnarray*}
\pi _{i} &=&{\frac{\partial \mathcal{L}_{k}}{\partial \dot{z}^{i}}}={\frac{-1%
}{{2\sqrt{-\mathcal{G}}}}}[2{\frac{\partial x_{0}^{a}}{\partial \dot{z}^{i}}}%
G_{ab}x_{1}^{b}(x_{0}^{a_{1}}G_{a_{1}b_{1}}x_{1}^{b_{1}})-2{\frac{\partial
x_{0}^{a}}{\partial \dot{z}^{i}}}%
G_{ab}x_{0}^{b}(x_{1}^{a_{1}}G_{a_{1}b_{1}}x_{1}^{b_{1}})] \\
&=&{\frac{\partial x_{0}^{a}}{\partial \dot{z}^{i}}}{\frac{\partial \mathcal{%
L}_{k}}{\partial x_{0}^{a}}}\equiv {\frac{\partial x_{0}^{a}}{\partial \dot{z%
}^{i}}}J_{a},
\end{eqnarray*}%
\begin{equation*}
J_{a}={\frac{-1}{\sqrt{-\mathcal{G}}}}%
[G_{ab}x_{1}^{b}(x_{0}^{a_{1}}G_{a_{1}b_{1}}x_{1}^{b_{1}})-G_{ab}x_{0}^{b}(x_{1}^{a_{1}}G_{a_{1}b_{1}}x_{1}^{b_{1}})]
\end{equation*}

\begin{equation*}
\pi _{i}\dot{z}^{i}={\frac{-1}{\sqrt{-\mathcal{G}}}}\dot{z}^{i}{\frac{%
\partial x_{0}^{a}}{\partial \dot{z}^{i}}}%
[G_{ab}x_{1}^{b}(x_{0}^{a_{1}}G_{a_{1}b_{1}}x_{1}^{b_{1}})-G_{ab}x_{0}^{b}(x_{1}^{a_{1}}G_{a_{1}b_{1}}x_{1}^{b_{1}})]=%
\dot{z}^{i}{\frac{\partial x_{0}^{a}}{\partial \dot{z}^{i}}}J_{a}.
\end{equation*}

Since $\dot{z}^{i}{\frac{\partial x_{0}^{a}}{\partial \dot{z}^{i}}}=\alpha
_{i}^{a}\dot{z}^{i}=x_{0}^{a}-x_{00}^{a}$, so we can derive

\begin{eqnarray*}
\pi _{i}\dot{z}^{i} &=&{\frac{-1}{\sqrt{-\mathcal{G}}}}%
[(x_{0}^{a}-x_{00}^{a})G_{ab}x_{1}^{b}(x_{0}^{a_{1}}G_{a_{1}b_{1}}x_{1}^{b_{1}})-(x_{0}^{a}-x_{00}^{a})G_{ab}x_{0}^{b}(x_{1}^{a_{1}}G_{a_{1}b_{1}}x_{1}^{b_{1}})]
\\
&=&-\sqrt{-\mathcal{G}}+{\frac{1}{\sqrt{-\mathcal{G}}}}%
[x_{00}^{a}G_{ab}x_{1}^{b}(x_{0}^{a_{1}}G_{a_{1}b_{1}}x_{1}^{b_{1}})-x_{00}^{a}G_{ab}x_{0}^{b}(x_{1}^{a_{1}}G_{a_{1}b_{1}}x_{1}^{b_{1}})]
\\
&=&\mathcal{L}_{k}+{\frac{1}{\sqrt{-\mathcal{G}}}}%
[x_{00}^{a}G_{ab}x_{1}^{b}(x_{0}^{a_{1}}G_{a_{1}b_{1}}x_{1}^{b_{1}})-x_{00}^{a}G_{ab}x_{0}^{b}(x_{1}^{a_{1}}G_{a_{1}b_{1}}x_{1}^{b_{1}})],
\end{eqnarray*}
and
\begin{eqnarray}
\mathcal{\tilde{H}} &=&\pi _{i}\dot{z}^{i}-\mathcal{L}_{k}=x_{00}^{a}G_{ab}{%
\frac{%
[x_{1}^{b}(x_{0}^{a_{1}}G_{a_{1}b_{1}}x_{1}^{b_{1}})-x_{0}^{b}(x_{1}^{a_{1}}G_{a_{1}b_{1}}x_{1}^{b_{1}})]%
}{\sqrt{-\mathcal{G}}}}  \notag \\
&=&-x_{00}^{a}{\frac{\partial \mathcal{L}_{k}}{\partial x_{0}^{a}}}%
=-x_{00}^{a}J_{a}.  \label{A3}
\end{eqnarray}

We can check the identities
\begin{equation*}
J_{a}x_{1}^{a}={\frac{\partial \mathcal{L}_{k}}{\partial x_{0}^{a}}}%
x_{1}^{a}={\frac{-1}{\sqrt{-\mathcal{G}}}}%
[x_{1}^{a}G_{ab}x_{1}^{b}(x_{0}^{a_{1}}G_{a_{1}b_{1}}x_{1}^{b_{1}})-x_{1}^{a}G_{ab}x_{0}^{b}(x_{1}^{a_{1}}G_{a_{1}b_{1}}x_{1}^{b_{1}})]=0,
\end{equation*}
and
\begin{eqnarray*}
J_{a}(G^{-1})_{ab}J_{b} &=&{\frac{\partial \mathcal{L}_{k}}{\partial
x_{0}^{a}}}G_{ab}^{-1}{\frac{\partial \mathcal{L}_{k}}{\partial x_{0}^{b}}}={%
\frac{1}{\sqrt{-\mathcal{G}}}}%
[G_{ab}x_{1}^{b}(x_{0}^{a_{1}}G_{a_{1}b_{1}}x_{1}^{b_{1}})-G_{ab}x_{0}^{b}(x_{1}^{a_{1}}G_{a_{1}b_{1}}x_{1}^{b_{1}})]
\\
&&\times G_{ac}^{-1}{\frac{1}{\sqrt{-\mathcal{G}}}}%
[G_{cd}x_{1}^{d}(x_{0}^{a_{1}}G_{a_{1}b_{1}}x_{1}^{b_{1}})-G_{cd}x_{0}^{d}(x_{1}^{a_{1}}G_{a_{1}b_{1}}x_{1}^{b_{1}})]
\\
&=&-x_{1}^{a}G_{ab}x_{1}^{b}.
\end{eqnarray*}

These two equations may help us to solve $\frac{\partial \mathcal{L}_{k}}{%
\partial x_{0}^{+}}$ and $\frac{\partial \mathcal{L}_{k}}{\partial x_{0}^{9}}
$ as functions of other $\frac{\partial \mathcal{L}_{k}}{\partial x_{0}^{a}}$%
.

Another identity
\begin{eqnarray*}
{\frac{\partial \mathcal{L}_{k}}{\partial x_{0}^{a}}}x_{0}^{a} &=&{\frac{-1}{%
\sqrt{-\mathcal{G}}}}%
[x_{0}^{a}G_{ab}x_{1}^{b}(x_{0}^{a_{1}}G_{a_{1}b_{1}}x_{1}^{b_{1}})-x_{0}^{a}G_{ab}x_{0}^{b}(x_{1}^{a_{1}}G_{a_{1}b_{1}}x_{1}^{b_{1}})]
\\
&=&{\frac{-1}{\sqrt{-\mathcal{G}}}}%
[(x_{0}Gx_{1})(x_{0}Gx_{1})-(x_{0}Gx_{0})(x_{1}Gx_{1})] \\
&=&-\sqrt{(x_{0}Gx_{1})^{2}-(x_{0}Gx_{0})(x_{1}Gx_{1})}=-\sqrt{-\mathcal{G}}=%
\mathcal{L}_{k},
\end{eqnarray*}%
can be used in deriving (\ref{A3}).

The momenta $\pi _{i}$ are
\begin{equation*}
\pi _{i}={\frac{\partial x_{0}^{a}}{\partial \dot{z}_{i}}}J_{a},
\end{equation*}%
where $x_{\mu }^{a}\equiv L_{\mu }^{a}.$

The nonzero $\frac{\partial x_{0}^{a}}{\partial \dot{z}^{i}}$ are
\begin{equation*}
{\frac{\partial x_{0}^{-}}{\partial \dot{x}^{-}}}=e^{\phi },{\frac{\partial
x_{0}^{x}}{\partial \dot{x}}}=e^{\phi },{\frac{\partial x_{0}^{\bar{x}}}{%
\partial \dot{\bar{x}}}=}e^{\phi },{\frac{\partial x_{0}^{D}}{\partial \dot{x%
}^{D}}}={\frac{\partial x_{0}^{D}}{\partial \dot{\phi}}}=1,{\frac{\partial
x_{0}^{a^{\prime }}}{\partial \dot{y}^{a^{\prime }}}=}u^{a^{\prime
}},a^{\prime }=5,6,7,8.
\end{equation*}%
One has
\begin{equation*}
\pi _{-}=e^{\phi }J_{-},\pi _{x}=e^{\phi }J_{x},\pi _{\bar{x}}=e^{\phi }J_{%
\bar{x}},\pi _{D}=J_{D},\pi _{a^{\prime }}=u^{a^{\prime }}J_{a^{\prime }},
\end{equation*}%
and
\begin{equation}
J_{-}=e^{-\phi }\pi _{-},J_{x}=e^{-\phi }\pi _{x},J_{\bar{x}}=e^{-\phi }\pi
_{\bar{x}},J_{D}=\pi _{D},J_{a^{\prime }}={\frac{1}{u^{a^{\prime }}}}\pi
_{a^{\prime }}.  \label{A-4}
\end{equation}%
Noting from (\ref{8-20})(\ref{3-3}),
\begin{eqnarray}
L^{+} &=&e^{\phi }dx^{+},  \notag \\
L^{-} &=&e^{\phi }[dx^{-}-{\frac{i}{2}(}\theta ^{i}d\theta _{i}+\theta
_{i}d\theta ^{i})]-{\frac{1}{2}}e^{-\phi }[{\frac{1}{4}}({\eta ^{2}}%
)^{2}dx^{+}+{\frac{i}{2}(}\eta ^{i}d\eta _{i}+\eta _{i}d\eta ^{i})],  \notag
\\
L^{x} &=&e^{\phi }dx,\text{ \ \ \ \ \ \ \ \ \ \ \ }L^{\bar{x}}=e^{\phi }d{%
\bar{x}},  \notag \\
L^{D} &=&d\phi ,\text{ \ \ \ \ \ \ \ \ \ \ \ }L^{A^{\prime }}=u^{A^{\prime
}}dx^{A^{\prime }}+v^{A^{\prime }}dx^{+},\text{ \ }  \label{A-5}
\end{eqnarray}%
where ${\eta ^{2}=}\eta ^{i}\eta _{i}$ and defining%
\begin{eqnarray}
b_{0\theta } &=&-{\frac{i}{2}(}\theta ^{i}\dot{\theta}_{i}+\theta _{i}\dot{%
\theta}^{i}{)},b_{0\eta }={\frac{i}{2}(}\eta ^{i}\dot{\eta}_{i}+\eta _{i}%
\dot{\eta}^{i}{)},  \notag \\
b_{1\theta } &=&-{\frac{i}{2}(\theta ^{i}\theta _{i}^{\prime }+\theta _{i}{%
\theta }^{\prime i})},b_{1\eta }={\frac{i}{2}(}\eta ^{i}\eta _{i}^{\prime
}+\eta _{i}\eta ^{\prime i}{)},
\end{eqnarray}

from (\ref{23a}) and (\ref{A-5}) one obtains
\begin{eqnarray}
x_{0}^{+} &=&e^{\phi }\dot{x}^{+},\text{ \ \ \ }x_{0}^{-}=e^{\phi }(\dot{x}%
^{-}+b_{0\theta })-{\frac{1}{2}}e^{-\phi }[{\frac{1}{4}}({\eta }^{2})^{2}%
\dot{x}^{+}+b_{0\eta }],  \notag \\
x_{0}^{x} &=&e^{\phi }\dot{x},\text{ \ \ \ \ }x_{0}^{\bar{x}}=e^{\phi }\dot{%
\bar{x}},x_{0}^{D}=\dot{\phi},x_{0}^{A^{\prime }}=u^{A^{\prime }}\dot{y}%
^{A^{\prime }}+v^{A^{\prime }}\dot{x}^{+},  \notag \\
x_{1}^{+} &=&e^{\phi }x^{\prime +},\text{ \ \ \ }x_{1}^{-}=e^{\phi
}(x^{\prime -}+b_{1\theta })-{\frac{1}{2}}e^{-\phi }[{\frac{1}{4}}({\eta }%
^{2})^{2}x^{\prime +}+b_{1\eta }],  \notag \\
x_{1}^{x} &=&e^{\phi }x^{\prime },\text{ \ \ \ \ }x_{1}^{\bar{x}}=e^{\phi }{%
\ \bar{x}}^{\prime },x_{1}^{D}=\phi ^{\prime },x_{1}^{A^{\prime
}}=u^{A^{\prime }}y^{\prime A^{\prime }}+v^{A^{\prime }}x^{\prime +}.
\end{eqnarray}

For gauge $x^{+}=\tau ,y^{9}=\sigma $, we have
\begin{eqnarray}
\dot{x}^{+} &=&1,\text{ \ \ \ }x^{\prime +}=0,\text{ \ \ }\dot{y}^{9}=0,%
\text{ \ \ \ }y^{\prime 9}=1,  \notag \\
x_{0}^{A^{\prime }} &=&u^{A^{\prime }}\dot{y}^{A^{\prime }}+v^{A^{\prime }},%
\text{ \ \ \ \ \ }x_{1}^{A^{\prime }}=u^{A^{\prime }}y^{\prime A^{\prime }},
\notag \\
x_{0}^{+} &=&e^{\phi },\text{ \ \ \ \ \ \ \ \ \ \ \ \ \ \ }x_{1}^{+}=0 \\
x_{0}^{9} &=&v^{9},\text{ \ \ \ \ \ \ \ \ \ \ \ \ \ \ }x_{1}^{9}=u^{9}=1.
\end{eqnarray}

This implies
\begin{equation*}
x_{00}^{+}=e^{\phi },\text{ \ \ \ }x_{00}^{-}=e^{\phi }b_{0\theta }-{\frac{1%
}{2}}e^{-\phi }[b_{0\eta }+{\frac{1}{4}}(\eta ^{2})^{2}],\text{ \ \ \ }%
x_{00}^{A^{\prime }}=v^{A^{\prime }}.
\end{equation*}

The equation $J_{a}x_{1}^{a}=0$ gives
\begin{equation*}
J_{+}x_{1}^{+}+J_{-}x_{1}^{-}+\cdots +J_{8}x_{1}^{8}+J_{9}x_{1}^{9}=0.
\end{equation*}

So we can derive
\begin{eqnarray}
J_{9} &=&-({\frac{1}{x_{1}^{9}}})[J_{+}x_{1}^{+}+J_{-}x_{1}^{-}+\cdots
+J_{8}x_{1}^{8}]  \notag \\
&=&-\{\pi _{-}(x^{\prime -}+b_{1\theta }-{\frac{1}{2}}e^{-2\phi }b_{1\eta
})+\pi _{x}x^{\prime }+\pi _{\bar{x}}\bar{x}^{\prime }+\pi _{D}\phi ^{\prime
}+\sum_{a^{\prime }=5}^{8}\pi _{a^{\prime }}y^{\prime a^{\prime }}\}.
\end{eqnarray}

We have $G_{ab}^{-1}=G_{ab}$. Equation $J_{a}(G_{ab}^{-1}J_{b})=-x_{1}Gx_{1}$
gives
\begin{eqnarray}
&&2J_{+}J_{-}+2J_{x}J_{\bar{x}}+J_{D}^{2}+\sum_{a^{\prime
}=5}^{8}J_{a^{\prime }}^{2}+J_{9}^{2}  \notag \\
&=&-[2x_{1}^{+}x_{1}^{-}+2x_{1}^{x}x_{1}^{\bar{x}}+({x_{1}^{D}}%
)^{2}+\sum_{a^{\prime }=5}^{8}(x_{1}^{a^{\prime }})^{2}+(x_{1}^{9})^{2}] \\
&=&-[2e^{2\phi }x^{\prime }\bar{x}^{\prime }+\phi ^{\prime
2}+\sum_{a^{\prime }=5}^{8}(u^{a^{\prime }}x^{\prime a^{\prime }})^{2}+1],
\end{eqnarray}%
and
\begin{eqnarray}
J_{+} &=&{\frac{1}{{2J_{-}}}}\{-[2e^{2\phi }x^{\prime }\bar{x}^{\prime
}+\phi ^{\prime 2}+(u^{a^{\prime }})^{2}(x^{\prime a^{\prime
}})^{2}+1]-[2e^{-2\phi }\pi _{x}\pi _{\bar{x}}+\pi _{D}^{2}+\sum_{a^{\prime
}=5}^{8}(u^{a^{\prime }})^{-2}\pi _{a^{\prime }}^{2}  \notag \\
&&+(\pi _{-}(x^{\prime -}+b_{1\theta }-{\frac{1}{2}}e^{-2\phi }b_{1\eta
})+\pi _{x}x^{\prime }+\pi _{\bar{x}}\bar{x}^{\prime }+\pi _{D}\phi ^{\prime
}+\sum_{a^{\prime }=5}^{8}\pi _{a^{\prime }}x^{\prime a^{\prime }})^{2}]\}.
\end{eqnarray}

One has
\begin{equation*}
\tilde{\mathcal{H}}=-x_{00}^{a}J_{a},
\end{equation*}

\bigskip with
\begin{equation*}
x_{00}^{+}=e^{\phi },x_{00}^{-}=e^{\phi }b_{0\theta }-{\frac{1}{2}}e^{-\phi
}[b_{0\eta }+{\frac{1}{4}}(\eta ^{2})^{2}],x_{00}^{a^{\prime }}=v^{a^{\prime
}},x_{00}^{9}=v^{9},
\end{equation*}

giving
\begin{eqnarray}
\tilde{\mathcal{H}} &=&-J_{+}e^{\phi }-J_{-}[e^{\phi }b_{0\theta }-{\frac{1}{%
2}}e^{-\phi }[b_{0\eta }+{\frac{1}{4}}(\eta ^{2})^{2}]]-\sum_{a^{\prime
}=5}^{8}J_{a^{\prime }}v^{a^{\prime }}-J_{9}v^{9}  \notag \\
&=&{\frac{e^{2\phi }}{{2\pi _{-}}}}\{[2e^{2\phi }x^{\prime }\bar{x}^{\prime
}+{\phi ^{\prime }}^{2}+\sum_{a^{\prime }=5}^{8}(u^{a^{\prime
}})^{2}(y^{\prime a^{\prime }})^{2}+1]+[2e^{-2\phi }\pi _{x}\pi _{\bar{x}%
}+\pi _{D}^{2}+\sum_{a^{\prime }=5}^{8}(u^{a^{\prime }})^{-2}\pi _{a^{\prime
}}^{2}]  \notag \\
&&+[(\pi _{-}(x^{\prime -}+b_{1\theta }-{\frac{1}{2}}e^{-2\phi }b_{1\eta
})+\pi _{x}x^{\prime }+\pi _{\bar{x}}\bar{x}^{\prime }+\pi _{D}\phi ^{\prime
}+\sum_{a^{\prime }=5}^{8}\pi _{a^{\prime }}y^{\prime a^{\prime }}]^{2}\}
\notag \\
&&-e^{-\phi }\pi _{-}[e^{\phi }b_{0\theta }-{\frac{1}{2}}e^{-\phi }[b_{0\eta
}+{\frac{1}{4}}(\eta ^{2})^{2}]-\sum_{a^{\prime }=5}^{8}\pi _{a^{\prime }}{%
\frac{v^{a^{\prime }}}{{u^{a^{\prime }}}}}  \notag \\
&&+v^{9}[\pi _{-}(x^{\prime -}+b_{1\theta }-{\frac{1}{2}}e^{-2\phi }b_{1\eta
})+\pi _{x}x^{\prime }+\pi _{\bar{x}}\bar{x}^{\prime }+\pi _{D}\phi ^{\prime
}+\sum_{a^{\prime }=5}^{8}\pi _{a^{\prime }}y^{\prime a^{\prime }}]  \notag
\\
&=&{\frac{e^{2\phi }}{{2\pi _{-}}}}\{[2e^{2\phi }x^{\prime }\bar{x}^{\prime
}+{\phi ^{\prime }}^{2}+\sum_{a^{\prime }=5}^{8}(u^{a^{\prime
}})^{2}(y^{\prime a^{\prime }})^{2}+1]+[2e^{-2\phi }\pi _{x}\pi _{\bar{x}%
}+\pi _{D}^{2}+\sum_{a^{\prime }=5}^{8}(u^{a^{\prime }})^{-2}\pi _{a^{\prime
}}^{2}]  \notag \\
&&+[(\pi _{-}(x^{\prime -}-{\frac{i}{2}({\theta ^{i}\theta }^{\prime }{%
_{i}+\theta _{i}{\theta }^{\prime i}})}-{\frac{i}{4}}e^{-2\phi }{(\eta
^{i}\eta _{i}^{\prime }+\eta _{i}\eta ^{\prime i})})+\pi _{x}x^{\prime }+\pi
_{\bar{x}}\bar{x}^{\prime }+\pi _{D}\phi ^{\prime }  \notag \\
&&+\sum_{a^{\prime }=5}^{8}\pi _{a^{\prime }}y^{\prime a^{\prime
}}]^{2}\}+\pi _{-}[{\frac{i}{2}(\theta ^{i}\dot{\theta}_{i}+\theta _{i}\dot{%
\theta}^{i})}+{\frac{i}{4}e^{-2\phi }(\eta ^{i}\dot{\eta}_{i}+\eta _{i}\dot{%
\eta}^{i})}+{\frac{1}{8}}e^{-2\phi }(\eta ^{2})^{2}]  \notag \\
&&-\sum_{a^{\prime }=5}^{8}\pi _{a^{\prime }}{\frac{v^{a^{\prime }}}{{%
u^{a^{\prime }}}}}+v^{9}[\pi _{-}(x^{\prime -}-{\frac{i}{2}({\theta
^{i}\theta }^{\prime }{_{i}+\theta _{i}{\theta }^{\prime i}})}-{\frac{i}{4}}%
e^{-2\phi }{(\eta ^{i}\eta _{i}^{\prime }+\eta _{i}\eta ^{\prime i})})
\notag \\
&&+\pi _{x}x^{\prime }+\pi _{\bar{x}}\bar{x}^{\prime }+\pi _{D}\phi ^{\prime
}+\sum_{a^{\prime }=5}^{8}\pi _{a^{\prime }}y^{\prime a^{\prime }}].
\end{eqnarray}%
They are functions of coordinates for $\tilde{\mathcal{L}}_{k}$. Therefore
the partial Legendre transformed Lagrangian density is

\begin{equation*}
\tilde{\mathcal{L}}_{k}=\pi _{-}\dot{x}^{-}+\pi _{x}\dot{x}+\pi _{\bar{x}}%
\dot{\bar{x}}+\pi _{D}\dot{\phi}+\sum_{a^{\prime }=5}^{8}\pi _{a^{\prime }}%
\dot{y}^{a^{\prime }}-\tilde{\mathcal{H}}
\end{equation*}%
\begin{eqnarray}
&&=\pi _{-}\dot{x}^{-}+\pi _{x}\dot{x}+\pi _{\bar{x}}\dot{\bar{x}}+\pi _{D}%
\dot{\phi}+\sum_{a^{\prime }=5}^{8}\pi _{a^{\prime }}\dot{y}^{a^{\prime }}
\notag \\
&&-{\frac{e^{2\phi }}{{2\pi _{-}}}}\{[2e^{2\phi }x^{\prime }\bar{x}^{\prime
}+{\phi ^{\prime }}^{2}+\sum_{a^{\prime }=5}^{8}(u^{a^{\prime
}})^{2}(y^{\prime a^{\prime }})^{2}+1]+[2e^{-2\phi }\pi _{x}\pi _{\bar{x}%
}+\pi _{D}^{2}+\sum_{a^{\prime }=5}^{8}(u^{a^{\prime }})^{-2}\pi _{a^{\prime
}}^{2}]  \notag \\
&&+[(\pi _{-}(x^{\prime -}-{\frac{i}{2}({\theta ^{i}\theta }^{\prime }{%
_{i}+\theta _{i}{\theta }^{\prime i}})}-{\frac{i}{4}}e^{-2\phi }{(\eta
^{i}\eta _{i}^{\prime }+\eta _{i}\eta ^{\prime i})})+\pi _{x}x^{\prime }+\pi
_{\bar{x}}\bar{x}^{\prime }+\pi _{D}\phi ^{\prime }  \notag \\
&&+\sum_{a^{\prime }=5}^{8}\pi _{a^{\prime }}y^{\prime a^{\prime
}}]^{2}\}-\pi _{-}[{\frac{i}{2}(\theta ^{i}\dot{\theta}_{i}+\theta _{i}\dot{%
\theta}^{i})}+{\frac{i}{4}e^{-2\phi }(\eta ^{i}\dot{\eta}_{i}+\eta _{i}\dot{%
\eta}^{i})}+{\frac{1}{8}}e^{-2\phi }(\eta ^{2})^{2}]  \notag \\
&&+\sum_{a^{\prime }=5}^{8}\pi _{a^{\prime }}{\frac{v^{a^{\prime }}}{{%
u^{a^{\prime }}}}}-v^{9}[\pi _{-}(x^{\prime -}-{\frac{i}{2}({\theta
^{i}\theta }^{\prime }{_{i}+\theta _{i}{\theta }^{\prime i}})}+{\frac{i}{4}}%
e^{-2\phi }{(\eta ^{i}\eta _{i}^{\prime }+\eta _{i}\eta ^{\prime i})})
\notag \\
&&+\pi _{x}x^{\prime }+\pi _{\bar{x}}\bar{x}^{\prime }+\pi _{D}\phi ^{\prime
}+\pi _{a^{\prime }}y^{\prime a^{\prime }}].
\end{eqnarray}%
It satisfies ${\frac{\partial \tilde{\mathcal{L}}_{k}}{\partial \tau }}=0$
and is linear in velocities. In addition to that, the locality condition $%
\mathcal{A}_{ij}=0$ is satisfied.

The WZ term
\begin{eqnarray*}
\mathcal{L}_{WZ} &=&-{\frac{e^{\phi }}{\sqrt{2}}}\epsilon ^{\mu \nu
}\partial _{\mu }x^{+}[\tilde{\eta ^{i}}C_{ij}^{\prime }({\partial _{\nu }%
\tilde{\theta}^{j}}+i\partial _{\nu }x\tilde{\eta ^{j}})-\tilde{\eta ^{i}}%
C_{ij}^{\prime }({\partial _{\nu }\tilde{\theta}_{j}}-i\partial _{\nu }x%
\tilde{\eta _{j}})] \\
&=&-{\frac{e^{\phi }}{\sqrt{2}}}[\tilde{\eta ^{i}}C_{ij}^{\prime }(\tilde{%
\theta}^{\prime j}+ix^{\prime }\tilde{\eta ^{j}})-\tilde{\eta ^{i}}%
C_{ij}^{\prime }(\tilde{\theta}_{j}^{\prime }+ix^{\prime }\tilde{\eta _{j}}%
)],
\end{eqnarray*}%
and the total Lagrangian density
\begin{eqnarray}
\mathcal{\tilde{L}} &=&\tilde{\mathcal{L}}_{k}+\mathcal{L}_{WZ}  \notag \\
&=&\pi _{-}\dot{x}^{-}+\pi _{x}\dot{x}+\pi _{\bar{x}}\dot{\bar{x}}+\pi _{D}%
\dot{\phi}+\sum_{a^{\prime }=5}^{8}\pi _{a^{\prime }}\dot{y}^{a^{\prime }}
\notag \\
&&-\pi _{-}[{\frac{i}{2}(\theta ^{i}\dot{\theta}_{i}+\theta _{i}\dot{\theta}%
^{i})}+{\frac{i}{4}e^{-2\phi }(\eta ^{i}\dot{\eta}_{i}+\eta _{i}\dot{\eta}%
^{i})]-}\mathcal{H}  \notag \\
&=&f_{i}\dot{z}^{i}+f_{\alpha }\dot{z}^{\alpha }{-}\mathcal{H}.
\end{eqnarray}
also does.

We have the Hamiltonian density%
\begin{eqnarray*}
\mathcal{H} &\mathcal{=}&\mathcal{\tilde{H}}-\pi _{-}[{\frac{i}{2}(\theta
^{i}\dot{\theta}_{i}+\theta _{i}\dot{\theta}^{i})}+{\frac{i}{4}e^{-2\phi
}(\eta ^{i}\dot{\eta}_{i}+\eta _{i}\dot{\eta}^{i})]-}\mathcal{L}_{WZ} \\
&=&-x_{00}^{a}J_{a}-\pi _{-}[{\frac{i}{2}(\theta ^{i}\dot{\theta}_{i}+\theta
_{i}\dot{\theta}^{i})}+{\frac{i}{4}e^{-2\phi }(\eta ^{i}\dot{\eta}_{i}+\eta
_{i}\dot{\eta}^{i})]-}\mathcal{L}_{WZ}
\end{eqnarray*}%
\begin{eqnarray}
&=&-J_{+}e^{\phi }-J_{-}[e^{\phi }b_{0\theta }-{\frac{1}{2}}e^{-\phi
}[b_{0\eta }+{\frac{1}{4}}(\eta ^{2})^{2}]]-\sum_{a^{\prime
}=5}^{8}J_{a^{\prime }}v^{a^{\prime }}-J_{9}v^{9}  \notag \\
&&-\pi _{-}[{\frac{i}{2}(\theta ^{i}\dot{\theta}_{i}+\theta _{i}\dot{\theta}%
^{i})}+{\frac{i}{4}e^{-2\phi }(\eta ^{i}\dot{\eta}_{i}+\eta _{i}\dot{\eta}%
^{i})]-}\mathcal{L}_{WZ}  \notag \\
&=&-{\frac{\partial \mathcal{L}_{k}}{\partial \dot{x}^{+}}}-{\frac{\partial
\mathcal{L}_{WZ}}{\partial \dot{x}^{+}}}  \notag \\
&=&-\pi _{+}.
\end{eqnarray}

\section{Derivation for poisson bracket in field theory}

We first define $\frac{\delta }{\delta _{\epsilon }z_{i}(\sigma )}L[z]$. Let
$z_{i}(\sigma )$ be a function of $\sigma $, $\tilde{z}_{i}(\sigma )$ be
another function of $\sigma $, where%
\begin{equation}
\Delta _{\epsilon }z_{i}(\sigma )=\tilde{z}_{i}(\sigma )-z_{i}(\sigma
)=\delta _{\epsilon }(\sigma -\sigma ^{\prime })\eta _{i}(\sigma ^{\prime }),
\label{B1}
\end{equation}%
with $\int d\sigma \delta _{\epsilon }(\sigma -\sigma ^{\prime })=1$, $%
\delta _{\epsilon }(x)=\delta _{\epsilon }(-x),\delta _{\epsilon
}(x+L)=\delta _{\epsilon }(x),\frac{\partial ^{n}}{\partial \sigma ^{n}}%
\delta _{\epsilon }(\sigma -\sigma ^{\prime })=\delta _{\epsilon
}^{(n)}(\sigma -\sigma ^{\prime })$ exist. This is a smoothed change of $%
z_{i}(\sigma )$ centered at $\sigma ^{\prime }$.

One has%
\begin{equation*}
\underset{\epsilon \rightarrow 0}{\lim }\int d\sigma f(\sigma )\delta
_{\epsilon }(\sigma -\sigma ^{\prime })=f(\sigma ^{\prime }),
\end{equation*}%
for continuous $f(\sigma ).$

In (\ref{B1}), $\eta _{i}(\sigma ^{\prime })$ is an infinitesimal Grassmann
number or an ordinary infinitesimal number. Define%
\begin{eqnarray*}
\frac{\delta }{\delta _{\epsilon }z_{i}(\sigma ^{\prime })}L &=&\frac{1}{%
\eta _{i}(\sigma ^{\prime })}\{L[\tilde{z}]-L[z]\}, \\
\frac{\delta }{\delta z_{i}(\sigma ^{\prime })}L &=&\underset{\epsilon
\rightarrow 0}{\lim }\frac{\delta }{\delta _{\epsilon }z_{i}(\sigma ^{\prime
})}L.
\end{eqnarray*}

We then have

\begin{eqnarray}
\frac{\delta }{\delta _{\epsilon }z_{j}(\sigma ^{\prime })}\frac{\delta }{%
\delta _{\epsilon }z_{i}(\sigma )} &=&(-1)^{\hat{\imath}\hat{\jmath}}\frac{%
\delta }{\delta _{\epsilon }z_{i}(\sigma )}\frac{\delta }{\delta _{\epsilon
}z_{j}(\sigma ^{\prime })},  \notag \\
\frac{\delta }{\delta _{\epsilon }z_{i}(\sigma )}(AB) &=&(\frac{\delta A}{%
\delta _{\epsilon }z_{i}(\sigma )})B+(-1)^{\hat{\imath}\hat{a}}A(\frac{%
\delta B}{\delta _{\epsilon }z_{i}(\sigma )}),  \notag \\
\frac{\delta }{\delta _{\epsilon }z_{i}(\sigma )}(A+B) &=&\frac{\delta A}{%
\delta _{\epsilon }z_{i}(\sigma )}+\frac{\delta B}{\delta _{\epsilon
}z_{i}(\sigma )}.  \label{B0}
\end{eqnarray}%
where the indices $\hat{i},\hat{j},\hat{a}$ denote the Grassmann indices of $%
z_{i}(\sigma ),z_{j}(\sigma ^{\prime })$ and $A$ respectively. In text and
in this appendix, we use $(-1)^{ia}$ instead of $(-1)^{\hat{i}\hat{a}}$ if
no confusion.

For an action
\begin{equation*}
S=\int d\sigma d\tau \mathcal{L}(z,z^{\prime },\dot{z},\sigma ,\tau ),
\end{equation*}%
the field equation is
\begin{eqnarray*}
\delta S &=&0 \\
&=&\int d\sigma d\tau \delta z_{i}\{\frac{\partial \mathcal{L}}{\partial
z_{i}}-\partial _{\tau }(\frac{\partial \mathcal{L}}{\partial \dot{z}_{i}}%
)-\partial _{\sigma }(\frac{\partial \mathcal{L}}{\partial z_{i}^{\prime }}%
)\}+\text{surface terms ,}
\end{eqnarray*}%
giving%
\begin{equation*}
\frac{\partial \mathcal{L}}{\partial z_{i}(\sigma )}-\partial _{\tau }(\frac{%
\partial \mathcal{L}}{\partial \dot{z}_{i}(\sigma )})-\partial _{\sigma }(%
\frac{\partial \mathcal{L}}{\partial z_{i}^{\prime }(\sigma )})=0.
\end{equation*}%
We can alternatively write it as\
\begin{equation*}
\frac{\delta \mathcal{L}}{\delta _{\epsilon }z_{i}(\sigma )}-\partial _{\tau
}(\frac{\delta \mathcal{L}}{\delta _{\epsilon }\dot{z}_{i}(\sigma )})=0,
\end{equation*}%
if no boundary term.

For
\begin{equation*}
L\mathcal{(}z_{i},z^{\prime },\dot{z}_{i},\sigma ,\tau )=\int d\sigma
f_{i}(z(\sigma ),z^{\prime }(\sigma ))\dot{z}_{i}(\sigma )-\int d\sigma
g(z(\sigma ),z^{\prime }(\sigma )),
\end{equation*}%
the Lagrangian equation is
\begin{equation*}
\lbrack \frac{\partial f_{j}}{\partial z_{i}}-(-1)^{i+j+ij}\frac{\partial
f_{i}}{\partial z_{j}}-(\frac{\partial f_{j}}{\partial z_{i}^{\prime }}%
)^{\prime }]\dot{z}_{j}+[-\frac{\partial f_{j}}{\partial z_{i}^{\prime }}%
-(-1)^{i+j+ij}\frac{\partial f_{i}}{\partial z_{j}^{\prime }}]\dot{z}%
_{j}^{\prime }-\frac{\partial g}{\partial z_{i}}+(\frac{\partial g}{\partial
z_{i}^{\prime }})^{\prime }=0.
\end{equation*}%
With
\begin{eqnarray*}
\tilde{\omega}_{ij} &=&\frac{\partial f_{j}}{\partial z_{i}}-(-1)^{i+j+ij}%
\frac{\partial f_{i}}{\partial z_{j}}-(\frac{\partial f_{j}}{\partial
z_{i}^{\prime }})^{\prime }, \\
\mathcal{A}_{ij} &=&-\frac{\partial f_{j}}{\partial z_{i}^{\prime }}%
-(-1)^{i+j+ij}\frac{\partial f_{i}}{\partial z_{j}^{\prime }},
\end{eqnarray*}%
we can write it as%
\begin{equation*}
\tilde{\omega}_{ij}\dot{z}_{j}+\mathcal{A}_{ij}\dot{z}_{j}^{\prime }=\frac{%
\partial g}{\partial z_{i}}-(\frac{\partial g}{\partial z_{i}^{\prime }}%
)^{\prime }.
\end{equation*}%
Define%
\begin{equation*}
\hat{\omega}_{i(\sigma ),j(\sigma ^{\prime })}=\frac{\delta f_{j}(z(\sigma
^{\prime }),z^{\prime }(\sigma ^{\prime }))}{\delta _{\epsilon }z_{i}(\sigma
)}-(-1)^{i+j+ij}\frac{\delta f_{i}(z(\sigma ),z^{\prime }(\sigma ))}{\delta
_{\epsilon }z_{j}(\sigma ^{\prime })},
\end{equation*}%
one has
\begin{equation*}
J_{0}=\frac{\delta }{\delta _{\epsilon }z_{k}(\sigma _{k})}\hat{\omega}%
_{i(\sigma _{i}),j(\sigma _{j})}(-1)^{j^{2}+kj}+cyc(i,j,k)=0.
\end{equation*}%
We have

\begin{eqnarray*}
&&\int d\sigma ^{\prime }\hat{\omega}_{i(\sigma ),j(\sigma ^{\prime })}\dot{z%
}_{j}(\sigma ^{\prime }) \\
&=&\int d\sigma ^{\prime }[\delta _{\epsilon }(\sigma ^{\prime }-\sigma )%
\frac{\partial f_{j}}{\partial z_{i}}(\sigma ^{\prime })+\delta _{\epsilon
}^{\prime }(\sigma ^{\prime }-\sigma )\frac{\partial f_{j}}{\partial
z_{i}^{\prime }}(\sigma ^{\prime })]\dot{z}_{j}(\sigma ^{\prime }) \\
&&-\int d\sigma ^{\prime }(-1)^{i+j+ij}[\delta _{\epsilon }(\sigma -\sigma
^{\prime })\frac{\partial f_{i}}{\partial z_{j}}(\sigma )+\delta _{\epsilon
}^{\prime }(\sigma -\sigma ^{\prime })\frac{\partial f_{i}}{\partial
z_{j}^{\prime }}(\sigma )]\dot{z}_{j}(\sigma ^{\prime }) \\
&=&\int d\sigma ^{\prime }[\delta _{\epsilon }(\sigma ^{\prime }-\sigma )%
\frac{\partial f_{j}}{\partial z_{i}}(\sigma ^{\prime })-(-1)^{i+j+ij}\delta
_{\epsilon }(\sigma -\sigma ^{\prime })\frac{\partial f_{i}}{\partial z_{j}}%
(\sigma )-\delta _{\epsilon }(\sigma ^{\prime }-\sigma )(\frac{\partial f_{j}%
}{\partial z_{i}^{\prime }}(\sigma ^{\prime }))^{\prime }]\dot{z}_{j}(\sigma
^{\prime }) \\
&&+\int d\sigma ^{\prime }[-\delta _{\epsilon }(\sigma ^{\prime }-\sigma )%
\frac{\partial f_{j}}{\partial z_{i}^{\prime }}(\sigma ^{\prime
})-(-1)^{i+j+ij}\delta _{\epsilon }(\sigma -\sigma ^{\prime })\frac{\partial
f_{i}}{\partial z_{j}^{\prime }}(\sigma )]\dot{z}_{j}^{\prime },
\end{eqnarray*}%
giving
\begin{equation}
\underset{\epsilon \rightarrow 0}{\lim }\int d\sigma ^{\prime }\hat{\omega}%
_{i(\sigma ),j(\sigma ^{\prime })}\dot{z}_{j}(\sigma ^{\prime })=\frac{%
\partial g}{\partial z_{i}}-(\frac{\partial g}{\partial z_{i}^{\prime }}%
)^{\prime }.  \label{B2}
\end{equation}%
When $\mathcal{A}_{ij}=0,$ let the inverse matrix of $\tilde{\omega}$ be $%
\tilde{\Omega}$,%
\begin{equation*}
\tilde{\omega}_{ij}(\sigma )\tilde{\Omega}_{jk}(\sigma )=\delta _{ik},\tilde{%
\Omega}_{ij}(\sigma )\tilde{\omega}_{jk}(\sigma )=\delta _{ik}.
\end{equation*}%
One can prove%
\begin{equation}
\frac{\delta }{\delta _{\epsilon }z_{k}(\sigma _{k})}\tilde{\Omega}%
_{ij}(\sigma )=-\tilde{\Omega}_{il}(\sigma )\frac{\delta }{\delta _{\epsilon
}z_{k}(\sigma _{k})}\tilde{\omega}_{lm}(\sigma )\tilde{\Omega}_{mj}(\sigma
)(-1)^{(i+l)k}.  \label{B-a}
\end{equation}

Field equation (\ref{B2}) can be written as%
\begin{equation*}
\tilde{\omega}_{ij}(\sigma )\dot{z}_{j}(\sigma )=\underset{\epsilon
\rightarrow 0}{\lim }\int d\sigma ^{\prime }\frac{\delta }{\delta _{\epsilon
}z_{i}(\sigma )}g(\sigma ^{\prime }),
\end{equation*}%
and%
\begin{equation}
\dot{z}_{l}(\sigma )=\underset{\epsilon \rightarrow 0}{\lim }\int d\sigma
^{\prime }\tilde{\Omega}_{li}(\sigma )\frac{\delta }{\delta _{\epsilon
}z_{i}(\sigma )}g(\sigma ^{\prime }).  \label{B3}
\end{equation}%
Denote
\begin{equation*}
\hat{\Omega}_{i(\sigma _{i}),j(\sigma _{j})}=\tilde{\Omega}_{ij}(\sigma
)\delta _{\epsilon }(\sigma -\sigma ^{\prime }),
\end{equation*}%
and define the poisson bracket of functionals A and B as
\begin{equation*}
\{A,B\}=\underset{\epsilon \rightarrow 0}{\lim }\int d\sigma \int d\sigma
^{\prime }A\frac{\overleftarrow{\delta }}{\delta _{\epsilon }z_{i}(\sigma )}%
\hat{\Omega}_{i(\sigma ),j(\sigma ^{\prime })}\frac{\overrightarrow{\delta }B%
}{\delta _{\epsilon }z_{j}(\sigma ^{\prime })}.
\end{equation*}%
Equations (\ref{B3})\ is then
\begin{equation*}
\dot{z}_{i}(\sigma )=\{z_{i}(\sigma ),H\},
\end{equation*}%
where $H=\int d\sigma g(z(\sigma ),z^{\prime }(\sigma ))$.

We next study the key property of poisson bracket, the Jacobi identity. In
contrast with $J_{0},$
\begin{equation*}
J_{1}\equiv \frac{\delta }{\delta _{\epsilon }z_{k}(\sigma _{k})}[(-1)^{j+kj}%
\tilde{\omega}_{ij}(\sigma _{i})\delta _{\epsilon }(\sigma _{i}-\sigma
_{j})]+cyc(i,j,k),
\end{equation*}%
it is not identically zero however. The difference%
\begin{eqnarray*}
&&\frac{\delta }{\delta _{\epsilon }z_{k}(\sigma _{k})}\hat{\omega}%
_{i(\sigma ),j(\sigma ^{\prime })}-\frac{\delta }{\delta _{\epsilon
}z_{k}(\sigma _{k})}[\tilde{\omega}_{ij}(\sigma _{i})\delta _{\epsilon
}(\sigma _{i}-\sigma _{j})] \\
&=&-\delta _{\epsilon }(\sigma _{i}-\sigma _{j})[g_{ki}^{j}(\sigma
_{i})-g_{ki}^{j}(\sigma _{j})] \\
&&+\delta _{\epsilon }^{\prime }(\sigma _{i}-\sigma _{j})[G_{ki}^{j}(\sigma
_{i})-G_{ki}^{j}(\sigma _{j})] \\
&&+\delta _{\epsilon }(\sigma _{i}-\sigma _{j})\frac{\partial }{\partial
\sigma _{i}}G_{ki}^{j}(\sigma _{i}) \\
&\equiv &\Psi _{ijk}(\epsilon ,\sigma _{i},\sigma _{j},\sigma _{k}),
\end{eqnarray*}%
where
\begin{eqnarray*}
g_{ki}^{j}(\sigma ) &=&\delta _{\epsilon }(\sigma -\sigma _{k})\frac{%
\partial ^{2}f_{j}}{\partial z_{k}\partial z_{i}}(\sigma )+\delta _{\epsilon
}^{\prime }(\sigma -\sigma _{k})\frac{\partial ^{2}f_{j}}{\partial
z_{k}^{\prime }\partial z_{i}}(\sigma ), \\
G_{ki}^{j}(\sigma ) &=&\delta _{\epsilon }(\sigma -\sigma _{k})\frac{%
\partial ^{2}f_{j}}{\partial z_{k}\partial z_{i}^{\prime }}(\sigma )+\delta
_{\epsilon }^{\prime }(\sigma -\sigma _{k})\frac{\partial ^{2}f_{j}}{%
\partial z_{k}^{\prime }\partial z_{i}^{\prime }}(\sigma ),
\end{eqnarray*}%
has the property
\begin{equation}
\underset{\epsilon \rightarrow 0}{\lim }\int d\sigma _{i}d\sigma _{j}d\sigma
_{k}F_{i}(\epsilon ,\sigma _{i})F_{j}(\epsilon ,\sigma _{j})F_{k}(\epsilon
,\sigma _{k})\Psi _{ijk}(\epsilon ,\sigma _{i},\sigma _{j},\sigma _{k})=0,
\label{B5}
\end{equation}%
for smooth periodic functions $F_{i},F_{j},F_{k}$ with
\begin{equation*}
\underset{\epsilon \rightarrow 0}{\lim }F_{l}^{(n)}(\epsilon ,\sigma
)=F_{l}^{(n)}(\sigma ),l=i,j,k.
\end{equation*}%
one has
\begin{equation}
\underset{\epsilon \rightarrow 0}{\lim }\int d\sigma _{i}d\sigma _{j}d\sigma
_{k}\prod\limits_{l=i,j,k}F_{l}(\epsilon ,\sigma _{l})J_{1}=0.  \label{B6}
\end{equation}

Consider%
\begin{eqnarray*}
&&J(A,B,C) \\
&=&(-1)^{ca}\{A,\{B,C\}\}+(-1)^{ab}\{B,\{C,A\}\}+(-1)^{bc}\{C,\{A,B\}\} \\
&=&\underset{\epsilon \rightarrow 0}{\lim }\int d\sigma _{j}d\sigma
_{k}(-1)^{ac}(-1)^{j(b-1)}\{A,\frac{\delta B}{\delta _{\epsilon
}z_{j}(\sigma )}\tilde{\Omega}_{jk}(\sigma _{j})\delta _{\epsilon }(\sigma
_{j}-\sigma _{k})\frac{\delta C}{\delta _{\epsilon }z_{k}(\sigma _{k})}%
\}+cyc.(A,B,C)
\end{eqnarray*}
\begin{eqnarray}
&=&\underset{\epsilon ^{\prime }\rightarrow 0}{\lim }\int d\sigma
_{i}d\sigma _{l}\frac{\delta A}{\delta _{\epsilon ^{\prime }}z_{i}(\sigma
_{i})}\tilde{\Omega}_{il}(\sigma _{i})\delta _{\epsilon ^{\prime }}(\sigma
_{i}-\sigma _{l})\frac{\delta }{\delta _{\epsilon ^{\prime }}z_{l}(\sigma
_{l})}  \notag \\
&\times &\underset{\epsilon \rightarrow 0}{\lim }[\int d\sigma _{j}d\sigma
_{k}\frac{\delta B}{\delta _{\epsilon }z_{j}(\sigma )}\tilde{\Omega}%
_{jk}(\sigma _{j})\delta _{\epsilon }(\sigma _{j}-\sigma _{k})\frac{\delta C%
}{\delta _{\epsilon }z_{k}(\sigma _{k})}](-1)^{ac+j(b-1)+i(a-1)}+cyc(A,B,C)
\notag \\
&=&\underset{\epsilon ^{\prime }\rightarrow 0}{\lim }\underset{\epsilon
\rightarrow 0}{\lim }\int d\sigma _{i}d\sigma _{l}d\sigma _{j}d\sigma _{k}
\notag \\
&&\{(-1)^{ac+j(b-1)+i(a-1)}[\frac{\delta A}{\delta _{\epsilon ^{\prime
}}z_{i}(\sigma _{i})}\tilde{\Omega}_{il}(\sigma _{i})\delta _{\epsilon
^{\prime }}(\sigma _{i}-\sigma _{l})  \notag \\
&&\frac{\delta ^{2}B}{\delta _{\epsilon ^{\prime }}z_{l}(\sigma _{l})\delta
_{\epsilon }z_{j}(\sigma _{j})}\tilde{\Omega}_{jk}(\sigma _{j})\delta
_{\epsilon }(\sigma _{j}-\sigma _{k})\frac{\delta C}{\delta _{\epsilon
}z_{k}(\sigma _{k})}]  \notag \\
&&+(-1)^{ac+j(b-1)+i(a-1)+l(b+j)}[\frac{\delta A}{\delta _{\epsilon ^{\prime
}}z_{i}(\sigma _{i})}\tilde{\Omega}_{il}(\sigma _{i})\delta _{\epsilon
^{\prime }}(\sigma _{i}-\sigma _{l})  \notag \\
&&\frac{\delta B}{\delta _{\epsilon }z_{j}(\sigma _{j})}\frac{\delta }{%
\delta _{\epsilon ^{\prime }}z_{l}(\sigma _{l})}\tilde{\Omega}_{jk}(\sigma
_{j})\delta _{\epsilon }(\sigma _{j}-\sigma _{k})\frac{\delta C}{\delta
_{\epsilon }z_{k}(\sigma _{k})}]  \notag \\
&&+(-1)^{ac+j(b-1)+i(a-1)+l(b+k)}[\frac{\delta A}{\delta _{\epsilon ^{\prime
}}z_{i}(\sigma _{i})}\tilde{\Omega}_{il}(\sigma _{i})\delta _{\epsilon
^{\prime }}(\sigma _{i}-\sigma _{l})  \notag \\
&&\frac{\delta B}{\delta _{\epsilon }z_{j}(\sigma _{j})}\tilde{\Omega}%
_{jk}(\sigma _{j})\delta _{\epsilon }(\sigma _{j}-\sigma _{k})\frac{\delta
^{2}C}{\delta _{\epsilon ^{\prime }}z_{l}(\sigma _{l})\delta _{\epsilon
}z_{k}(\sigma _{k})}]\}  \notag \\
&&+cyc.(A,B,C)\equiv \alpha +\beta +\gamma .  \label{B4}
\end{eqnarray}%
This is a two fold limit. We first prove that the double limit $\underset{%
\epsilon \rightarrow 0,\epsilon ^{\prime }\rightarrow 0}{\lim }$ exists and
which does not depend on the manner of $\epsilon ,\epsilon ^{\prime
}\rightarrow 0$.

For this purpose, we use integration by parts to change all $\frac{\partial
}{\partial \sigma }\delta _{\epsilon ,\epsilon ^{\prime }}(\sigma -\sigma
^{\prime })$ to $-\delta _{\epsilon ,\epsilon ^{\prime }}(\sigma -\sigma
^{\prime })\frac{\partial }{\partial \sigma }$, in a properly chosen route.
Eventually we have an expression with only $\delta _{\epsilon ,\epsilon
^{\prime }}$ functions and local functions in the integration. The boundary
terms in integration by parts disappear because of the periodicity of local
functions and the property of the\ conserved quantities.

Then one sees that the double limit $\underset{\epsilon \rightarrow
0,\epsilon ^{\prime }\rightarrow 0}{\lim }$ is well-defined, it is
irrelevant of the manner $\epsilon ,\epsilon ^{\prime }\rightarrow 0$. We
have $\underset{\epsilon ^{\prime }\rightarrow 0}{\lim }[\underset{\epsilon
\rightarrow 0}{\lim }\{\cdots \}]=$ $\underset{\epsilon \rightarrow
0,\epsilon ^{\prime }\rightarrow 0}{\lim }\{\cdots \}=$ $\underset{\epsilon
=\epsilon ^{\prime }\rightarrow 0}{\lim }\{\cdots \},$if the first limit $%
\underset{\epsilon \rightarrow 0}{\lim }\{\cdots \}$ exists(which is easy to
check.), due to the multi limit theorem.

Since $\delta _{\epsilon }$ function is regular and $\frac{\delta }{\delta
_{\epsilon }}$operation is also regular, thus before limit, we can check the
validity of (\ref{B0}) and treat these terms just as in the finite
dimensional mechanics.

We then check $\alpha +\beta +\gamma $ in (\ref{B4}), and find when $%
\epsilon =\epsilon ^{\prime }$, $\alpha +\gamma =0$ because of the cyclic
permutation. The term $\underset{\epsilon \rightarrow 0}{\lim }\beta $ can
be converted to an integration of the form as l.h.s. of (\ref{B6}) by (\ref%
{B-a}) , and it is also zero due to the equation (\ref{B6}).

\end{document}